\documentclass[aps,twocolumn,reprint,superscriptaddress,groupedaddress,floatfix]{revtex4-1}
\usepackage{amsmath,amssymb,amsthm,amsfonts}
\usepackage{bbm}
\usepackage{graphicx}
\usepackage[hidelinks]{hyperref}

\newcommand{\avg}[1]{\langle #1 \rangle}
\newcommand{\D}{\mathcal{D}}
\renewcommand{\S}{\mathcal{S}}
\renewcommand{\d}{{\rm d}}
\newcommand{\s}{\sigma}
\newcommand{\up}{\uparrow}
\newcommand{\down}{\downarrow}

\newcommand{\nostar}{{\vphantom{*}}}
\newcommand{\mat}[1]{\underline{#1}}
\newcommand{\id}{\mat{\mathbbm{1}}}
\DeclareMathOperator{\Tr}{tr}

\begin{document}

\title{Charge Collective Modes in Correlated Electron Systems: Plasmons Beyond the Random Phase Approximation}

\author{Lo\"ic~Philoxene}

\author{Vu Hung~Dao}

\author{Raymond~Fr\'esard}
\email{raymond.fresard@ensicaen.fr}

\affiliation{Normandie Universit\'e, ENSICAEN, UNICAEN, CNRS, CRISMAT, 14000 Caen, France}

\date{\today}

\begin{abstract}
Elucidating the impact of strong electronic interactions on the collective excitations of metallic systems
has been of longstanding interest, mainly due to the inadequacy of the random phase approximation (RPA) in
the strongly correlated regime.
Here, we adopt our newly developed radial Kotliar and Ruckenstein slave boson representation to
analyze the charge excitation spectrum of a Hubbard model, extended with long range interactions.
Working on the face centered cubic lattice, at half filling, and in different coupling regimes ranging from
uncorrelated to the metal-to-insulator transition, we compare our results to conventional RPA as a benchmark.
We focus on the influence of the local and long range couplings on the particle-hole excitation continuum
and the plasmon and upper Hubbard band collective modes.
Beyond the weak coupling regime, we find numerous quantitative and even qualitative discrepancies between
our method and standard RPA.
Our work thus deepens the understanding of charge collective modes in correlated systems, and lays the
foundations for future
studies of a broad series of materials.
\end{abstract}

\maketitle

\paragraph*{Introduction.---}
In their seminal series of papers, Pines and Bohm pioneered the study of collective modes arising in
dynamical autocorrelation functions of the electron gas by introducing the random phase approximation
(RPA)~\cite{bohm1951,pines1951,bohm1953}.
Focusing on density fluctuations, they argued that their spectra may be split into two
components: i) an incoherent one associated with the random thermal motion of the individual
electrons, and ii) a plasma oscillation mode.
Having a classical analogue, the latter may be explained in simple terms,
and is broadly documented~\cite{anderson1994,schrieffer1999,giuliani2005}.
Nevertheless, quantum corrections to this classical picture were recently addressed~\cite{yan2016}.
Furthermore, a series of applications backing on its existence have been put forward,
ranging from nanophotonics~\cite{chang2007,lezec2007,akimov2007,noginov2009,oulton2009,cai2009,hryciw2010,schuller2010,ergin2010,mayer2011,botlasseva2011,soukoulis2011,huang2014,baev2015},
to energy conversion~\cite{atwater2010,linic2011,hou2011,lee2012,mukherjee2013}, and
even cancer treatment~\cite{oneal2004,li2012,swierczewska2012,smith2015,chavda2023}.

Since its introduction, it has been established that the RPA remains sensible in the weak coupling regime
only, and that it becomes unreliable as soon as the coupling strength becomes intermediate.
Nevertheless, it may still be applied as a flexible tool in the thermodynamic limit, and it indeed remains
broadly used, especially within quantum chemistry codes~\cite{lewis2019,li2023,scott2024}.
Besides, a series of calculations on model systems demonstrated qualitative failures of the approximation,
especially in the context of the celebrated one band Hubbard model.
In fact, key quantum collective phenomena entailed by the model, for example signatures of the upper Hubbard
band, are missing in the charge excitation spectra when computed within the RPA.
Multiple frameworks that try to overcome some of these shortcomings, and recover some of the missing features,
have thus been proposed~\cite{vanloon2014b,hafermann2014,guzzo2011,zhou2018,cudazzo2020,zinni2023}.

A broadly used approach to tackle correlated electrons is provided by Kotliar and Ruckenstein's slave boson
(KRSB) representation.
This versatile tool may be applied to a series of microscopic models, such as the Hubbard model~\cite{kotliar1986}
and its extensions~\cite{deeg1993,lhoutellier2015,riegler2023}.
It consists in introducing a doublet of pseudofermions, along with four bosons, that generate the Fock space on
each lattice site.
In the functional integral formulation, this results in a Lagrangian that is bi-linear in the fermionic fields,
although no Hubbard-Stratonovich decoupling is performed, thereby allowing for a description of electronic
interactions at arbitrary coupling strengths.
The reliability of the KRSB representation to the Hubbard model and its extensions has already been extensively
discussed (see, e.g., Paragraph II.C.1 in Ref.~\cite{philoxene2022} and references therein).
A recent study also put forward quantitative agreement between the charge and spin structure factors computed in
KRSB and resonant inelastic x-ray scattering data in electron-doped cuprates~\cite{riegler2023}.
Within this representation, calculations are amenable to the thermodynamic limit as well,
especially regarding the Mott metal-to-insulator transition~\footnote{
In particular, the paramagnetic saddle-point approximation of the KRSB representation is equivalent to the
Gutzwiller approximation, thereby describing the MIT at half filling. On the cubic lattice, the MIT critical coupling $U_c=1.33W$~\cite{lhoutellier2015}, where $W$ is the bare bandwidth, compares favorably with the DMFT result of $U_c=1.17W$~\cite{zitko2009}.
} (MIT), and does not suffer from a weak coupling limitation.
The resulting low energy spectra qualitatively differ from the RPA results, however~\cite{dao2017}.
They generically comprise a continuum, a zero-sound collective mode lying slightly above this continuum, and a
signature of the upper Hubbard band in the form of a mode that disperses about $\omega \sim U$ in the strong
coupling regime.
Below, we refer to the latter as the upper Hubbard band mode.
This mode may, in the intermediate coupling regime, hybridize with the zero-sound one~\cite{dao2017}.

In the past twenty years, the extended Hubbard model, entailing non-local density-density interactions, has seen an
upsurge of interest~\cite{aichhorn2004,davoudi2006,kagan2011,ayral2013,vanloon2014,kapcia2017,schuller2018,terletska2021,roig2022,philoxene2022,linner2023}
(see also~\cite{kundu2023} and references therein for a better overview).
It has also been studied early on within the KRSB representation, and phase diagrams have been computed
\cite{deeg1993}.
Below, we apply the radial gauge of the KRSB representation~\cite{fresard2001}, in which the non-local interaction
arises in a bi-linear form in terms of the boson fields.
It allows for a systematic evaluation of long range correlations between density fluctuations.
This formalism has recently been revisited and validated through exact calculations~\cite{dao2020,dao2024}.

The purpose of the present Letter is to compute the experimentally accessible energy loss spectrum of the extended
Hubbard model, via the calculation of the dynamical dielectric function, which itself depends on the charge
autocorrelation function.
Special focus is made on the interplay of the plasmon mode (driven by the non-local Coulomb interaction),
the upper Hubbard band (driven by the local interaction), and the low energy particle-hole excitation continuum.

\paragraph*{Model and methods.---}
The Hamiltonian for the Hubbard model, extended by a long range Coulomb interaction, may be written as
\begin{align}\label{hamiltonian}
\mathcal{H} =&
\sum_{i\ne j,\sigma} t_{ij} \left( c^{\dagger}_{\sigma,i} c^{\vphantom{\dagger}}_{\sigma,j} + {\rm h.c.} \right)
+ U \sum_{i} n_{\uparrow,i} n_{\downarrow,i} \notag\\
&+ \frac{1}{2} \sum_{i\ne j} V_{ij} \left(2-\sum_{\sigma}n_{\sigma,i}\right) \left(2-\sum_{\sigma}n_{\sigma,j}\right) ,
\end{align}
where $c_{i,\sigma}$ ($\sigma=\uparrow,\downarrow$) is the canonical electron annihilation operator,
$n_{i,\sigma}$ is the associated electron number operator, $t_{ij}=-t$ if $i$ and $j$ are nearest neighbors,
and $t_{ij}=0$ otherwise, and we set $\hbar=1$ here and throughout.
Here, $t$ is the hopping amplitude, $U$ is the Hubbard coupling, and
$V_{ij}=V a / |{\bf r}_i - {\bf r}_j|$ is the non-local Coulomb interaction, where $a$ is
the lattice spacing, and $V$ is an effective coupling parameter.
In the last term, the long range interaction has been chosen to couple to the hole densities $2-n$
for later convenience.
It is, however, equivalent to the representation in terms of electron densities $n$,
as they differ by an overall shift in energy, only.

We work in the grand canonical ensemble, and employ the radial gauge of the KRSB representation.
In this paragraph, we give an outline of the formalism behind this radial KRSB representation, and
refer to the Appendix~\ref{appA} for a detailed derivation and technical discussions.
In the original KRSB framework, 
one introduces a set of four auxiliary bosons $e$, $p_{\sigma}$, and $d$ (associated to unoccupied,
singly occupied with spin projection $\sigma$, and doubly occupied atomic states, respectively), as 
well as a doublet of pseudofermions $f_\sigma$ at each lattice site.
Within the functional integral formalism,
the canonical electron fields are mapped to a product of slave boson and pseudofermion fields as
\begin{align}
c_{\sigma,i}(\tau) \rightarrow z_{\sigma,i}(\tau) f_{\sigma,i}(\tau) ,
\end{align}
where the fields $z_{\sigma,i}(\tau)$ are functions of the boson fields (omitting the imaginary-time
variable $\tau$),
\begin{align}\label{eq:z}
z_{\sigma,i} =
e^*_{i} Y^{\vphantom{*}}_{\sigma,i} p^{\vphantom{*}}_{\sigma,i} + p^*_{-\sigma,i} Y^{\vphantom{*}}_{\sigma,i} d^{\vphantom{*}}_{i} ,
\end{align}
where
\begin{align}
Y_{\sigma,i} = \left[(e^*_{i}e^{\vphantom{*}}_{i} + p^*_{-\sigma,i}p^{\vphantom{*}}_{-\sigma,i})
(1 - p^*_{-\sigma,i}p^{\vphantom{*}}_{-\sigma,i} - e^*_{i}e^{\vphantom{*}}_{i})\right]^{-1/2} .
\end{align}
This representation is invariant under local ${\rm U(1) \times U(1) \times U(1)}$ gauge
transformations, allowing for the phase of three of the boson fields to be gauged away
\cite{jolicoeur1991,fresard1992,bang1992}.
The boson fields deprived of their phase degree of freedom are coined radial slave bosons.
Being real-valued, the radial slave boson fields are free from Bose condensation.
Their expectation values are
generically finite and can be well approximated in the thermodynamic limit via the 
saddle-point approximation. Corrections to the latter may be 
obtained when evaluating the Gaussian fluctuations~\cite{dao2017},
and the correspondence between this more precise evaluation and the time-dependent 
Gutzwiller approach~\cite{bunemann2013} could recently be achieved---though by means of
an extension in the formulation of the latter~\cite{noatschk2020}.
In the Cartesian gauge, the non-local interaction term reads
$\frac12\sum_{i\ne j}V_{ij}(2-p_{\uparrow,i}^2-p_{\downarrow,i}^2-2d^*_i d_i)(2-p_{\uparrow,j}^2-p_{\downarrow,j}^2-2d^*_j d_j)$.
It is quartic in the bosonic fields, and can therefore not be integrated exactly.
In contrast, in the radial gauge, one may re-write it as
$\frac12\sum_{i\ne j}V_{ij} ( 2 R_{e,i} + R_{\uparrow,i} + R_{\downarrow,i} ) ( 2 R_{e,j} + R_{\uparrow,j} + R_{\downarrow,j} )$,
after having made use of the constraints to eliminate the $d$-boson
(see Appendix~\ref{appA} and Ref.~\cite{dao2024}).
By doing so, both local and non-local interaction terms enter the action
as quadratic terms in the radial slave bosons.
Further details on the radial gauge are provided in Appendix~\ref{appA}.

\begin{figure*}
\centering
\includegraphics[width=.98\textwidth]{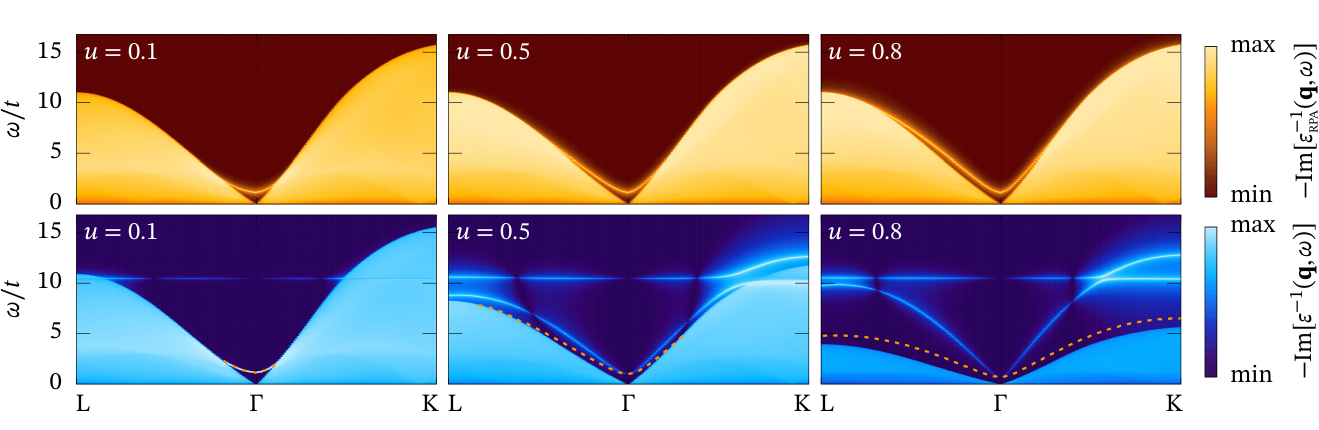}
\caption{Zero temperature RPA (top row) and radial KRSB (bottom row) energy loss spectra $-{\rm Im}[\varepsilon^{-1}({\bf q},\omega)]$, in dependence on ${\bf q}$ along ${L-\Gamma-K}$.
Parameters: $v=0.1$ and $u=0.1$, $0.5$, and $0.8$, from left to right.
The dashed lines in the bottom row denote the dispersion of the plasmon mode obtained by inserting the renormalized Lindhard function $\Pi_0$ in the expression of the RPA dielectric function Eq.~(\ref{eq:rpa})}
\label{fig:spec}
\end{figure*}

The charge excitation spectrum of the model can be analyzed through the evaluation of the
loss function $-{\rm Im}[\varepsilon^{-1}({\bf q},\omega)]$, which is experimentally accessible by
electron energy loss spectroscopy or resonant inelastic x-ray scattering measurements.

To that end, we compute the inverse dielectric function as
\begin{align}\label{eq:krsb}
\varepsilon^{-1}(q)=1 - \left( \frac{U}{2} + V_{\bf q} \right) \chi_c(q) ,
\end{align}
where $V_{\bf q}=4\pi a V/|{\bf q}|^2$, $q\equiv({\bf q},{\rm i}\omega_n$),
$\omega_n\equiv2\pi n k_B T$, $k_B$ is the Boltzmann constant, and
$T$ is the temperature. The density-density correlation function
\begin{align}
\chi_c(q) =& ~\avg{\delta n(-q) \delta n(q)} \notag\\
=& ~4 d^2 \avg{\delta d'(-q) \delta d'(q)}
- 2 d \avg{\delta d'(-q) \delta R_{e}(q)} \notag\\
&+ \avg{\delta R_{e}(-q) \delta R_{e}(q)} ,
\end{align}
is calculated by taking into account Gaussian fluctuations of the boson
fields around the paramagnetic saddle-point solution~\cite{dao2017}.
As detailed in the Appendix~\ref{appA}, we obtain analytical expressions for $\chi_c(q)$
and $\varepsilon^{-1}(q)$. The associated dynamical response functions
are then evaluated on the real frequency axis with the substitution
${\rm i} \omega_n \rightarrow \omega + {\rm i} 0^{+}$.
The advantage of our procedure over numerical methods, lies
in the obtained analytical expressions, which allow for an unambiguous
analytical continuation, and can be physically interpreted in some
limiting regimes.

The paramagnetic saddle-point of the KRSB representation has already been extensively studied in the
literature~\cite{kotliar1986,vollhardt1987,deeg1993,lhoutellier2015,dao2017}.
In radial gauge, the study of the saddle-point remains identical.

In this Letter, we focus on the face-centered-cubic (fcc) lattice as a representative example of three
dimensional systems, since the simple cubic structure is scarcely realized.
In this case, the bare dispersion is
$
t_{\bf k}=-4t \left(
\cos\frac{k_x}{2}\cos\frac{k_y}{2}
+ \cos\frac{k_y}{2}\cos\frac{k_z}{2}
+ \cos\frac{k_z}{2}\cos\frac{k_x}{2}
\right)
$,
where we set $a=1$.
As a proof of principle, we consider the half band-filling ($n=1$) case, which hosts the MIT,
thereby allowing us to unravel the impact of strong electron correlations on the loss
function.
In this context, the critical coupling of the Mott transition is
$U_c=-8\xi_{\bf 0}$,
where $\xi_{\bf 0}$ denotes the average bare kinetic energy.
We find $\xi_{\bf 0}\simeq-0.16W$, where $W=16t$ is the bare bandwidth.
This yields $U_c=1.31W$, which, due to the large coordination number of
the fcc lattice, also compares favorably with the large coordination
limit of dynamical mean-field theory (DMFT): $U_c=1.47W$~\cite{bulla1999}.

\paragraph*{Results.---}
In the standard Hartree-Fock RPA (HF+RPA) framework, the density-density correlation function is computed as a series of
particle-hole bubble diagrams for non-interacting electrons, linked with bare interaction vertices
$U/2+V_{\bf q}$. 
Under such approximations, the dynamical dielectric function reads~\cite{giuliani2005}
\begin{align}\label{eq:rpa}
\varepsilon_{\rm RPA}({\bf q},\omega)=1+\left(\frac{U}{2}+V_{\bf q}\right)\Pi_0^{(0)}({\bf q},\omega) ,
\end{align}
where $\Pi_0^{(0)}({\bf q},\omega)$ is the Lindhard function for the non-interacting system.
Due to its perturbative essence, we cannot expect standard HF+RPA procedure to yield reasonable results
in the strong coupling regime (see~\cite{dao2017} and Appendix~\ref{appC} for an assessment of some of the key
features missing in the RPA treatment that are incorporated in the Cartesian and radial KRSB formalisms, respectively).
However, we use it as a benchmark to highlight strong correlation effects when comparing
it with the radial KRSB representation for values of $U$ and/or $V$ approaching $U_c$.
In the following, we use the dimensionless coupling parameters $u=U/U_c$ and $v=V/U_c$.

Let us now address representative examples of the energy loss spectra computed with Eq.~(\ref{eq:rpa})
and with Eq.~(\ref{eq:krsb}).
We fix the value of $v=0.1$ and investigate values of $u=0.1$, $0.5$, and $0.8$ at half filling
and zero temperature.
We also focus on values of ${\bf q}$ along the representative symmetry lines ${L-\Gamma-K}$,
with $L=(\pi,\pi,\pi)$, $\Gamma=(0,0,0)$, and $K=(\frac{3\pi}{2},\frac{3\pi}{2},0)$,
for the wavevector dependence.
On the face centered cubic lattice, the nearest neighbor distance is smallest (largest)
along the ${\Gamma-K}$ (${\Gamma-L}$) direction.
The computed spectra are presented in Fig.~\ref{fig:spec}.
They generically comprise a low energy particle-hole excitation continuum.
In RPA, this continuum is insensitive to the value of $u$, and it disperses from
$\omega(\Gamma)=0$, up to $\omega(K) \simeq 16t$.
In our radial KRSB calculations, however, the continuum strongly depends on the value of $u$.
Indeed, it is gradually narrowed by increasing the Hubbard coupling, with a
maximum of its dispersion at $\omega(K) \simeq 16t$ for $u=0.1$, in contrast to
$\omega(K) \simeq 6t$, only, for $u=0.8$.
This owes to the fact that the Lindhard function for the quasiparticles
$\Pi_0({\bf q},\omega)$ and the non-interacting Lindhard function $\Pi^{(0)}_0({\bf q},\omega)$
are related via renormalization.
For the considered paramagnetic phase, this reads (see Appendix~\ref{appB} for a discussion)
\begin{align}\label{pi0}
\Pi^{(0)}_0\left({\bf q},\omega\right) = z_0^2 \Pi_0({\bf q},z_0^2 \omega) ,
\end{align}
where, $z_0^2=\avg{z_{\sigma,i}^\dagger z_{\sigma,i}}$ is the inverse mass renormalization factor,
appearing as $z_0^2 t_{\bf k}$ in the dispersion relation of the quasiparticles.
One may thus explicitly see the decrease of the continuum's bandwidth, as the quasiparticle
residue $z_0^2=1-u^2$ approaches zero when $u$ approaches unity~\cite{dao2017}.

\begin{figure}
\centering
\includegraphics[width=.38\textwidth]{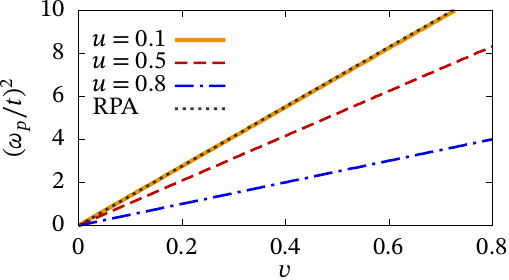}
\caption{Square of the plasma frequency $\omega_p^2$ in dependence on the strength of the effective Coulomb coupling $v$.
Parameters: $u=0.1$, $0.5$, and $0.8$.
The plasma frequency obtained in standard RPA is also shown.}
\label{fig:wpv}
\end{figure}

The RPA response features a single collective mode: the plasmon, which
establishes at large wavelengths, above the continuum.
When increasing ${\bf q}$, it enters the particle-hole continuum, and thus quickly becomes suppressed
by Landau damping.
For larger values of $u$, the mode gets overdamped at larger energy and wavevector,
especially in the ${L}$ direction, along which it disperses more.
Its gap at ${\bf q}=\Gamma$, the plasma frequency $\omega_p$, remains unchanged and in fact depends on $v$, only.
This can be understood from the expression
\begin{align}
\omega_p \simeq \sqrt{-\frac{V\xi_{\bf 0}}{6}} ,
\end{align}
obtained by a
large wavelength expansion of the RPA dielectric function (derived in Appendix~\ref{appA}).
For $v=0.1$ (i.e. $V=2.1t$), this yields $\omega_p \simeq t$, which coincides with the gap shown in the
top row of Fig.~\ref{fig:spec}.
Furthermore, we note that the plasma frequency computed with this expression does not depend on the
value of $u$.
In fact, $u$ first enters the dispersion of the plasmon mode as a contribution of order $|{\bf q}|^2$.
This additionally corroborates the observation that the plasmon mode disperses more for larger values
of the Hubbard coupling.
In the bottom row of Fig.~\ref{fig:spec}, we see that the radial KRSB spectra possess two well-defined
collective modes.
Firstly, we observe the plasmon mode, similarly to the RPA.
At weak coupling $u=0.1$, the plasmon collective mode is also present at large wavelengths, only, as it
enters the particle-hole continuum at approximately the same values of ${\bf q}$ and $\omega$ as
in RPA.
At larger couplings $u=0.5$ and $0.8$, though, the renormalization of the particle-hole continuum,
along with the greater dispersion of the plasmon mode induced by $u$, allows for the latter to remain
well-defined in a broader range of wavelengths.
One also sees that the plasmon mode obtained by inserting the renormalized polarizability
Eq.~(\ref{pi0}) in the RPA formula for the dielectric function correctly accounts for the
value of the plasma frequency at arbitrary coupling. At finite wavelengths, however, this
naive attempt to account for strong correlations within the RPA response noticeably deviates
from the KRSB plasmon.
Secondly, an additional collective mode establishes in the radial KRSB spectra.
This mode, with a much larger gap at $\Gamma$ of about $\omega_{\rm UHB} \simeq 10t$ for every
values of $u$, corresponds to the aforementioned upper Hubbard band mode.
Similarly to the plasmon mode, at weak coupling, it enters the particle-hole continuum at finite
${\bf q}$, inside which it quickly decays via Landau damping.
It, however, disperses much less than the plasmon mode, with a bandwidth of at most one for $u=0.8$.

In this two-modes picture, one might expect a level-crossing at finite ${\bf q}$ between the two branches,
at a given point of the parameter space.
Yet, no point of exact degeneracy could be found, but either no-crossing or anticrossings between
both modes, as depicted in the center and right panels of the bottom row in Fig.~\ref{fig:spec}.
Nonetheless, we observe multiple anticrossings with near-degeneracy, as can be seen for example close
to ${K}/2$ and energies around $\omega \simeq 10t$ for $u=0.8$.
Close to these anticrossings, the upper Hubbard band mode and the plasmon mode strongly hybridize,
and the excitations share both characters.

Fig.~\ref{fig:wpv} presents the $v$ dependence of the radial KRSB plasma frequency squared $\omega_p^2$, for
values of $u=0.1$, $0.5$, and $0.8$.
The RPA result is also shown for comparison, and one can see that in the weak
coupling regime, the radial KRSB formalism correctly reproduces the RPA plasma frequency, as expected.
Another expected property of the square of the plasma frequency is that it should scale with $v$, as
we have $\omega_p^2 \simeq 9.2v$ at weak coupling, which is also realized.
However, the deviation of the plasma frequency from this analytical expression is seen to increase
with $u$.
This indicates that strong correlation effects, arising when the local coupling becomes sufficiently
large, cause a softening of the plasmon mode, by opposition to the RPA picture in
which the plasmon mode is barely affected.

\begin{figure}
\centering
\includegraphics[width=.42\textwidth]{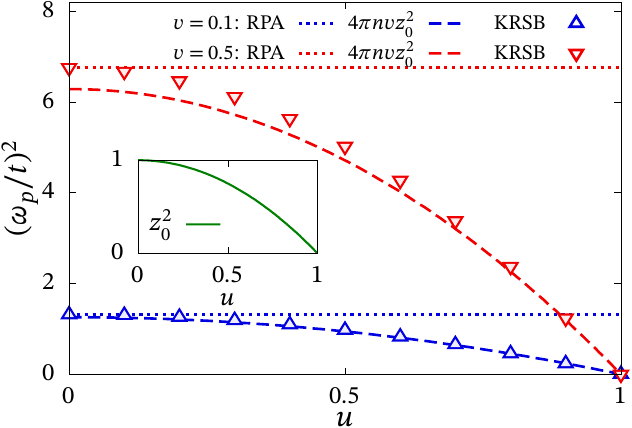}
\caption{
Square of the radial KRSB plasma frequency $\omega_p^2$ in dependence on the local coupling $u$.
The RPA results, and the plasma frequency obtained by using the classical expression (see text)
with $m^*=1/z_0^2$, are also shown.
Inset: Quasiparticle residue $z_0^2$ in dependence on $u$.
Parameters: $v=0.1$ and $0.5$.
}
\label{fig:wpu}
\end{figure}

The $u$ dependence of the radial KRSB plasma frequency squared is shown in Fig.~\ref{fig:wpu}, for
representative values of $v=0.1$ and $0.5$.
The (constant) RPA result, for the same values of $v$, is also presented for comparison.
We see that the plasma frequency decreases as the Hubbard coupling increases, ranging from the RPA value
for $u=0$, to zero at the onset of the Mott transition.
This can be qualitatively understood by considering the classical expression for the plasma frequency,
$\omega_p=\sqrt{4\pi ne^2/m^*}$, with $n$ the electron density, $e$ the electron charge,
and $m^*$ its effective mass.
Recalling that, for an unscreened Coulomb interaction $V=e^2$, and that the band
mass is given by the renormalization factor via $z_0^2t \sim 1/m^*$, we see that the
plasma frequency should decrease along with $z_0^2$ when the Hubbard coupling is increased.
This is better depicted in the inset of Fig.~\ref{fig:wpu}, in which $\omega_p^2$ is shown as a function
of the quasiparticle residue $z_0^2$.
We clearly see that at the onset of the MIT, at which the effective mass
diverges, $\omega_p$ drops to zero.
This vanishing of the plasma frequency at the onset of the Mott transition re-emphasizes the connection
between the coherent and collective nature of the plasmon mode.

\paragraph*{Summary and conclusion.---}
In summary, we have proposed a theoretical framework for the computation of
charge excitation spectra in the presence of long range Coulomb interactions,
and in the full range of correlation regimes.
We evidenced quantitative and qualitative discrepancies between our results
and standard RPA. In particular, we emphasized the influence of strong local
correlations on the plasmon collective mode, showing the possibility for the
plasmon to propagate undamped in broader ranges of wavelengths at strong
coupling, and recovering the expected dependence of the plasma frequency on
the renormalized mass of the electrons. At the onset of the Mott transition,
we found the plasma frequency to vanish along with the quasiparticle residue.
Regions of strong hybridization between the plasmon and upper Hubbard band
collective modes have also been unraveled.
The benefits of the obtained analytical formula for the loss function are
twofold. First, it allows for a straightforward analytical continuation,
thereby yielding sharply defined collective modes. Second, simple
approximations may be recovered in some limiting cases, providing
easy access to physical quantities.
As the computational cost is similar to that of standard RPA, our method,
and possible future generalizations, may be employed to refine the
incorporation of strong interactions in studies of correlated systems.

\begin{acknowledgments}
The authors gratefully thank Thilo Kopp for insightful discussions.
This work was supported by R\'egion Normandie through the ECOH project. 
Financial support provided by the ANR LISBON (ANR-20-CE05-0022-01) project is gratefully acknowledged.
\end{acknowledgments}

\appendix

\begin{widetext}
\section{Derivation of the radial KRSB density autocorrelation function} \label{appA}
\subsection{Expression of the radial KRSB action functional}
To circumvent the notorious failures of standard perturbation theory in interaction regimes
where the coupling scale approaches the characteristic electrons hopping energy scale $t$,
slave boson techniques introduce auxiliary fields in terms of which the interaction terms
may be written as bilinear terms. In the context of the extended Hubbard model, a particular
choice of such representation is that of Kotliar and Ruckenstein~\cite{kotliar1986}.
According to the exhaustive presentation provided in Ref.~\cite{dao2024},
the KRSB representation involves a doublet of fermionic fields $\{f_{\up,i},f_{\down,i}\}$, together
with four bosonic fields $\{e_i,p_{\up,i},p_{\down,i},d_i\}$ (omitting the time variables), which
are tied to empty, singly occupied (with spin projection $\s\in\{\up,\down\}$), and doubly occupied
lattice sites. In the radial gauge, the first three boson fields possess an amplitude degree of
freedom, only, labeled $R_{e,i}$ and $R_{\s,i}$. The grand canonical partition function then reads
\begin{align}
\mathcal{Z} =&\lim_{\nu\to 0}\lim_{N\to \infty}\lim_{\eta\to 0^+}\left[
\prod_{n=1}^N \prod_{i=1}^L \left(
\int_{-\eta}^\infty \d R_{e,i,n} \d R_{\up,i,n} \d R_{\down,i,n}
\int_{-\infty}^\infty \frac{\epsilon\d\alpha_{i,n}}{2\pi}
\frac{\epsilon\d\beta_{\up,i,n}}{2\pi}
\frac{\epsilon\d\beta_{\down,i,n}}{2\pi}
\frac{\d d'_{i,n} \d d''_{i,n}}{\pi}
\int \prod_\sigma \d f_{\sigma,i,n} \d f^*_{\sigma,i,n}
\right) \right. \notag\\
& \left.\vphantom{prod_i^L \int_a^a \frac{d}{d}} \times {\rm e}^{-(\S_f[f^*,f,\psi]+\S_b[\psi])} \right] ,
\end{align}
where $\epsilon=\beta/N$, and the vector
$\psi = (R_e, d^*, d, R_\up, R_\down, \beta_\up, \beta_\down, \alpha)$
gathers the boson fields, as well as the time-dependent fields $\alpha$ and $\beta_\s$.
The latter are enforcing the constraints
\begin{align}
&R_{e,i} + R_{\up,i} + R_{\down,i} + d_i'^2 + d_i''^2 = 1 , \\
&R_{\s,i} + d_i'^2 + d_i''^2 = f_{\s,i}^* f_{\s,i}^\nostar ,
\end{align}
respectively. The action functional entails a purely bosonic contribution
\begin{align}
\S_b[\psi] = \epsilon \sum_{n=1}^N \sum_{i=1}^L &\left\{
\epsilon^{-1}d^*_{i,n}\left(d_{i,n}-{\rm e}^{-\epsilon(U+{\rm i}\tilde{\alpha}_{i,n}-{\rm i}\beta_{\up,i,n}-{\rm i}\beta_{\down,i,n})}d_{i,n-1}\right)
+ \frac{1}{2} \sum_{j\ne i}\sum_{\sigma,\sigma'} V_{ij} (R_{\sigma,i,n}+R_{e,i,n})(R_{\sigma',j,n}+R_{e,j,n}) \right. \notag\\
&\left.
+{\rm i}\tilde{\alpha}_{i,n}\left(R_{e,i,n}+R_{\up,i,n}+R_{\down,i,n}-1\right)
-{\rm i}\sum_\sigma \beta_{\sigma,i,n} R_{\sigma,i,n} 
\right\},
\end{align}
and a mixed fermionic-bosonic contribution
\begin{align}
\S_f[f^*,f,\psi]=\sum_{n=1}^N \sum_{i=1}^L \sum_\sigma
f^*_{\sigma,i,n} \left(f_{\sigma,i,n}-{\rm e}^{-\epsilon({\rm i}\beta_{\sigma,i,n}-\mu_0)}f_{\sigma,i,n-1}
+ \epsilon\sum_{j\ne i} \check{z}_{\sigma,i,n} z_{\sigma,j,n-1} t_{ij} f_{\sigma,j,n-1} 
\right) .
\end{align}
Here, $\tilde{\alpha}_{i,n}=\alpha_{i,n}-{\rm i}\lambda_0$, with some regulator $\lambda_0>0$ ensuring
convergence of the integrals, and where the notation $\check{z}_{\sigma,i,n}$ has been
introduced to stress that this field is not the complex conjugate of $z_{\sigma,i,n}$,
but rather
\begin{align}
&z_{\sigma,i,n}=\frac{\sqrt{R_{e,i,n+1}R_{\sigma,i,n}}+\sqrt{R_{-\sigma,i,n+1}}d_{i,n}}{\sqrt{R_{e,i,n+1}+R_{-\sigma,i,n+1}-{\rm i}\nu}\sqrt{1-R_{e,i,n}-R_{-\sigma,i,n}+{\rm i}\nu}} ,\\
&\check{z}_{\sigma,i,n}=\frac{\sqrt{R_{\sigma,i,n+1}R_{e,i,n}}+d^*_{i,n}\sqrt{R_{-\sigma,i,n}}}{\sqrt{1-R_{e,i,n+1}-R_{-\sigma,i,n+1}-{\rm i}\nu}\sqrt{R_{e,i,n}+R_{-\sigma,i,n}+{\rm i}\nu}} .
\end{align}

Expanding the exponentials appearing in the action functional to leading order in $\epsilon$,
and using the continuous imaginary-time notation, these expressions may conveniently be rewritten
as
\begin{align}
\mathcal{Z} = \int \D[f^*,f,\psi] {\rm e}^{-(\S_f[f^*,f,\psi]+\S_b[\psi])} ,
\end{align}
with 
\begin{align}
\S_b[\psi] =& \int_0^{1/T}  \d\tau \sum_{i} \Big[
U ( d'^2_{i} + d''^2_{i} )
+ \frac{1}{2} \sum_{j\ne i} \sum_{\s,\s'} V_{ij} ( R_{\s,i} + R_{e,i} ) ( R_{\s',j} + R_{e,j} ) \notag\\
&+ {\rm i}\alpha_i ( R_{e,i} + d'^2_{i} + d''^2_{i} + R_{\up,i} + R_{\down,i} - 1 )
- {\rm i}\sum_\s \beta_{\s,i} ( R_{\s,i} + d'^2_{i} + d''^2_{i} )
+ {\rm i} ( d'_{i} \partial_\tau d''_{i} - d''_{i} \partial_\tau d'_{i} )
\Big] ,
\end{align}
and
\begin{align}
\S_f[f^*,f,\psi] = 
\int_0^{1/T}\d\tau \sum_{i,j} \sum_\s f^*_{\s,i} \left[
(\partial_\tau - \mu_0 + {\rm i}\beta_{\s,i}) \delta_{i,j} + z^*_{\s,i} z^\nostar_{\s,j} t_{ij}
\right] f^\nostar_{\s,j} ,
\end{align}
where the time labels $\tau=\lim_{N\to\infty} \epsilon$ have been omitted for clarity,
and $d'_i$ and $d''_i$ denote the real and imaginary parts of the $d_i$ boson, respectively.

One could then demand proofs of the correctness of such a representation. In the presence
of non-local interactions, this has been, most recently, provided by an explicit calculation
of the partition function and correlation functions of the two-site cluster.
Even though it acts as a toy integrable model, this limit of the extended Hubbard model has
been shown to harbor all the technical hurdles of larger system sizes~\cite{dao2024}.

\subsection{Expansion of the action functional around the paramagnetic saddle-point}

\subsubsection{Paramagnetic saddle-point}
At the paramagnetic saddle-point, the slave boson amplitudes take the expectation values
\begin{align}
\psi^{\rm SP} = (E, d, 0, P, P, \beta_0, \beta_0, \alpha) ,
\end{align}
the values of which are determined in dependence on the model parameters by solving the
saddle-point equation~\cite{kotliar1986,fresard1992,deeg1993,zimmermann1997,lhoutellier2015,dao2017}
\begin{align}\label{spe}
\frac{(1-x^2)x^4}{x^4-\delta^2} = \frac{U}{U_0} ,
\end{align}
where $\delta=1-n$ is the hole doping in a half-filled band, and $U_0$ is a coupling scale which,
at half filling ($\delta=0$), corresponds to the critical coupling of the interaction-driven Mott
transition.
At arbitrary doping, $U_0$ reads
\begin{align}
U_0=-\frac{8~\xi_{\bf 0}}{1-\delta^2} ,
\end{align}
with the semi-renormalized kinetic energy
\begin{align}
\xi_{\bf 0} = \frac{2}{L} \sum_{\bf k} n_F(E_{\bf k}) t_{\bf k} ,
\end{align}
where
\begin{align}
t_{\bf k} = -4t \left(
\cos\frac{k_x}{2} \cos\frac{k_y}{2}
+ \cos\frac{k_y}{2} \cos\frac{k_z}{2}
+ \cos\frac{k_z}{2} \cos\frac{k_x}{2}
\right) ,
\end{align}
is the bare dispersion of the fcc lattice, and
\begin{align}
E_{\bf k} = z_0^2 t_{\bf k} - (\mu - \beta_0) ,
\end{align}
is the quasiparticle dispersion, where $z_0 = \avg{z_{\sigma,i}}$.
Note that the value of $V$ does not enter Eq.~(\ref{spe}), hence the saddle-point
values of the $e$, $p_\s$, and $d$ slave boson amplitudes do not depend on $V$.
It turns out that the effect of $V$ is simply to induce a shift in the
values of the saddle-point amplitudes of $\alpha$ and $\beta_0$
\begin{align}
\left.\alpha\right|_{V} &= \left.\alpha\right|_{V=0} - 2(2-n)V_{\bf 0} , \\
\left.\beta_0\right|_{V} &= \left.\beta_0\right|_{V=0} - (2-n)V_{\bf 0} ,
\end{align}
that nevertheless leaves Eq.~(\ref{spe}) unaffected.
Here, $V_{\bf 0} = V_{{\bf q}=\Gamma}$.

\subsubsection{Gaussian fluctuations around the saddle-point}
In order to investigate the excitations of the paramagnetic ground state of the system,
we begin by expanding the action functional to second order in the field fluctuations
\begin{align}
\delta\psi_{\mu,i}(\tau) = \psi_{\mu,i}(\tau) - \psi^{\rm SP}_\mu ,
\end{align}
about the paramagnetic saddle-point.
This procedure yields contributions of increasing order in the field fluctuations,
from which we only retain the zero-th to second orders,
\begin{align}
\S[f^*,f,\psi] \simeq \S^{(0)}[f^*,f] + \S^{(1)}[f^*,f,\delta\psi] + \S^{(2)}[f^*,f,\delta\psi] .
\end{align}
Once again, we split the different orders of the contributions in bosonic $\S_b$
and mixed $\S_f$ sectors.
The bosonic sector then contains the contributions
\begin{align}
\S_b^{(0)} = \frac{L}{T} \left[
U d^2 + \frac{1}{2} V_{\bf 0} (2-n)^2 - 2 \beta_0 (P+d^2) + \alpha (E+2P+d^2-1)
\right] ,
\end{align}
\begin{align}
\S_b^{(1)}[\delta\psi] = \sqrt{\frac{L}{T}} &\left\{ \vphantom{\sum_\s}
[\alpha+2V_{\bf 0}(2-n)] \delta R_e(0)
+ 2d[\alpha-2\beta_0+U] \delta d'(0) \right.
+ [\alpha-\beta_0+V_{\bf 0}(2-n)] \sum_\s \delta R_\s(0) \notag\\
&-(P+d^2) \sum_\s \delta \beta_\s(0)
\left.+(E+2P+d^2-1) \delta \alpha(0) \vphantom{\sum_\s} \right\} ,
\end{align}
\begin{align}
\S_b^{(2)}[\delta\psi] = \sum_q \sum_{\mu,\nu} \delta\psi_\mu(-q) D^{-1}_{b,\mu\nu}(q) \delta\psi_\nu(q) ,
\end{align}
where $L$ is the number of lattice sites, $q=({\bf q},{\rm i}\omega_n)$, with $\omega_n$ a bosonic (i.e. even)
Matsubara frequency. We also introduced the $8\times8$ matrix
\begin{align}
\renewcommand{\arraystretch}{1.5}
\mat{D}_b^{-1}(q) =
\begin{pmatrix}
2V_{\bf q} &0 &0 &V_{\bf q} &V_{\bf q} &0 &0 &\frac{1}{2} \\
0 &\alpha-2\beta_0+U &\omega_n &0 &0 &-d &-d &d \\
0 &-\omega_n &\alpha-2\beta_0+U &0 &0 &0 &0 &0 \\
V_{\bf q} &0 &0 &\frac{1}{2}V_{\bf q} &\frac{1}{2}V_{\bf q} &-\frac{1}{2} &0 &\frac{1}{2} \\
V_{\bf q} &0 &0 &\frac{1}{2}V_{\bf q} &\frac{1}{2}V_{\bf q} &0 &-\frac{1}{2} &\frac{1}{2} \\
0 &-d &0 &-\frac{1}{2} &0 &0 &0 &0 \\
0 &-d &0 &0 &-\frac{1}{2} &0 &0 &0 \\
\frac{1}{2} &d &0 &\frac{1}{2} &\frac{1}{2} &0 &0 &0
\end{pmatrix} .
\end{align}
In the fermionic sector, we have to expand the $z$-factors in powers of the slave boson
fluctuations,
\begin{align}
z_{\s,i}(\tau)
&\simeq
z_0
+ \sum_\mu \left.\frac{\partial z_{\s,i}(\tau)}{\partial\psi_{\mu,i}(\tau)}\right|_{\psi=\psi^{\rm SP}} \delta\psi_{\mu,i}(\tau)
+ \frac{1}{2} \sum_{\mu,\nu} \left.\frac{\partial^2 z_{\s,i}(\tau)}{\partial\psi_{\mu,i}(\tau)\psi_{\nu,i}(\tau)}\right|_{\psi=\psi^{\rm SP}} \delta\psi_{\mu,i}(\tau) \delta\psi_{\nu,i}(\tau) \notag\\
&= z_0 + \sum_\mu Z_{\s,\mu}^{(1)} \delta\psi_{\mu,i}(\tau) + \frac{1}{2} \sum_{\mu,\nu} Z_{\s,\mu\nu}^{(2)} \delta\psi_{\mu,i}(\tau) \delta\psi_{\nu,i}(\tau) ,
\end{align}
where the derivatives of $z_\sigma$ are denoted
\begin{align}
&Z_{\s,\mu}^{(1)} = \left.\frac{\partial z_{\s,i}(\tau)}{\partial\psi_{\mu,i}(\tau)}\right|_{\psi=\psi^{\rm SP}} , \\
&Z_{\s,\mu\nu}^{(2)} = \left.\frac{\partial^2 z_{\s,i}(\tau)}{\partial\psi_{\mu,i}(\tau)\psi_{\nu,i}(\tau)}\right|_{\psi=\psi^{\rm SP}} .
\end{align}
Doing so, we find that the contributions, up to second order in the slave boson fields fluctuations,
sum up to
\begin{align}\label{dyn:eqsf}
\S_f[f^*,f,\delta\psi] \simeq
\sum_\s \sum_{k,k'} f_\s^*(k) [ -G_0^{-1}(k,k') + H_\s(k,k') ] f_\s^\nostar(k') ,
\end{align}
where $k=({\bf k},{\rm i}\omega_\ell)$, with $\omega_\ell$ a fermionic Matsubara
frequency, and
\begin{align}
&G_0^{-1}(k,k') = [{\rm i}\omega_\ell + (\mu-\beta_0) - z_0^2 t_{\bf k}] \delta_{k,k'} , \\
&H_\s(k,k') =
\sqrt{\frac{T}{L}} \Big\{
\delta\beta_\s(k-k') + z_0 \sum_\mu [\bar{Z}_{\s,\mu}^{(1)} t_{\bf k'} + Z_{\s,\mu}^{(1)} t_{\bf k} ] \delta\psi_\mu(k-k')
\Big\} \notag\\
&\hspace{4em}+ \frac{T}{L} \sum_q \sum_{\mu,\nu} \delta\psi_\mu(k-k'-q) \Big\{
\frac{z_0}{2} [ \bar{Z}_{\s,\mu\nu}^{(2)} t_{\bf k'} + Z_{\s,\mu\nu}^{(2)} t_{\bf k} ]
+ \bar{Z}_{\s,\mu}^{(1)} Z_{\s,\nu}^{(1)} t_{\bf k'+q}
\Big\} \delta\psi_\nu(q) ,
\end{align}
where $\bar{Z}_{\s,\mu}^{(1)}=(Z_{\s,\mu}^{(1)})^*$.
Using the integration rules for Gaussian integrals over Grassmann fields,
we explicitly integrate the partition function over the pseudofermions
\begin{align}\label{dyn:eqzf}
\mathcal{Z}_f &= \exp\left\{ \sum_\s \Tr\left[ \ln(-\mat{G}_0^{-1}+\mat{H}_\s) \right] \right\} \notag\\
&= \det(\mat{G}_0^{-1})^2 \exp\left\{ \sum_\s \Tr\left[ \ln(\id-\mat{G}_0\mat{H}_\s) \right] \right\} \notag\\
&\simeq \det(\mat{G}_0^{-1})^2 \exp\left\{ - \sum_\s \Tr\left[ \mat{G}_0\mat{H}_\s + \frac{1}{2}\mat{G}_0\mat{H}_\s\mat{G}_0\mat{H}_\s \right] \right\} \notag\\
&= \det(\mat{G}_0^{-1})^2 \exp\left\{ - \sum_q \sum_{\mu,\nu} \delta\psi_\mu(-q) D_{f,\mu\nu}^{-1}(q) \delta\psi_\nu(q) \right\} ,
\end{align}
where the trace is taken over the wavevectors and Matsubara frequencies.
For the first term, this yields
\begin{align}
\Tr(\mat{G}_0 \mat{H}_\s)
=& \sum_{k_1,k_2} G_0(k_1,k_2) H_\s(k_2,k_1) \notag\\
=& \sum_k G_0(k) H_\s(k,k) \notag\\
=& \sum_k G_0(k) \Big\{
\sqrt{\frac{T}{L}} \Big[
\delta\beta_\s(0) + 2 z_0 \sum_\mu {\rm Re}(Z_{\s,\mu}^{(1)}) t_{\bf k} \delta\psi_\mu(0)
\Big] \notag\\
&\hspace{4em}+ \frac{T}{L} \sum_q \sum_{\mu,\nu} \delta\psi_\mu(-q) \Big[
z_0 {\rm Re}(Z_{\s,\mu\nu}^{(2)}) t_{\bf k}
+ \bar{Z}_{\s,\mu}^{(1)} Z_{\s,\nu}^{(1)} t_{\bf k+q}
\Big] \delta\psi_\nu(q) \Big\} , \notag\\
=& -\frac{1}{2} \sum_q \sum_{\mu,\nu} \delta\psi_\mu(-q) \Big[
z_0 {\rm Re}(Z_{\s,\mu\nu}^{(2)}) \xi_{\bf 0}
+ \bar{Z}_{\s,\mu}^{(1)} Z_{\s,\nu}^{(1)} \xi_{\bf q}
\Big] \delta\psi_\nu(q) .
\end{align}
In the last line, we used the fact that the terms of first order in the fields fluctuations are
compensated by the bosonic part at the saddle-point, and we defined
\begin{align}
\xi_{\bf q} = -\frac{2T}{L} \sum_k G_0(k)\ t_{\bf k+q}
= \frac{2}{L} \sum_{\bf k} n_F(E_{\bf k}) t_{{\bf k}+{\bf q}}.
\end{align}
For the second term, we have, to second order in the boson fields fluctuations,
\begin{align}
\Tr(\frac{1}{2} \mat{G}_0 \mat{H}_\s \mat{G}_0 \mat{H}_\s)
=& \frac{1}{2} \sum_{k_1,\cdots,k_4} G_0(k_1,k_2) H_\s(k_2,k_3) G_0(k_3,k_4) H_\s(k_4,k_1) \notag\\
=& \frac{T}{2L} \sum_{k_1,k_2} G_0(k_1) G_0(k_2) \Big\{
\delta\beta_\s(k_1-k_2) \delta\beta_\s(k_2-k_1) \notag\\
&+\delta\beta_\s(k_1-k_2) z_0 \sum_\mu [\bar{Z}_{\s,\mu}^{(1)} t_{\bf k_1} + Z_{\s,\mu}^{(1)} t_{\bf k_2}] \delta\psi_\mu(k_2-k_1) + {\rm h.c.} \notag\\
&+\!\!\sum_{\mu,\nu} \delta\psi_\mu(k_1-k_2) z_0^2 [\bar{Z}_{\s,\mu}^{(1)} t_{\bf k_2} \!+\! Z_{\s,\mu}^{(1)} t_{\bf k_1}] [\bar{Z}_{\s,\nu}^{(1)} t_{\bf k_1} \!+\! Z_{\s,\nu}^{(1)} t_{\bf k_2}]  \delta\psi_\nu(k_2-k_1)
\Big\} \notag\\
=& \frac{T}{2L} \sum_q \sum_{\mu,\nu} \delta\psi_\mu(-q) \sum_k G_0(k) G_0(k+q) \Big\{
(\delta_{\mu\nu,66}\delta_{\s,\up} + \delta_{\mu\nu,77}\delta_{\s,\down}) \notag\\
&+ 2 z_0 [\bar{Z}_{\s,\nu}^{(1)} t_{\bf k} + Z_{\s,\nu}^{(1)} t_{\bf k+q}] (\delta_{\mu,6}\delta_{\s,\up}+\delta_{\mu,7}\delta_{\s,\down}) + {\rm h.c.} \notag\\
&+ z_0^2 [2{\rm Re}(Z_{\s,\mu}^{(1)} Z_{\s,\nu}^{(1)}) t_{\bf k} t_{\bf k+q} + \bar{Z}_{\s,\mu}^{(1)} Z_{\s,\nu}^{(1)} t_{\bf k+q}^2 + Z_{\s,\mu}^{(1)} \bar{Z}_{\s,\nu}^{(1)} t_{\bf k}^2]
\Big\} \delta\psi_\nu(q) \notag\\
=& -\frac{1}{2}\sum_q \sum_{\mu,\nu} \delta\psi_\mu(-q) \Big\{
\frac{1}{2} X_{00}(q) (\delta_{\mu\nu,66}\delta_{\s,\up} + \delta_{\mu\nu,77}\delta_{\s,\down}) \notag\\
&+ z_0 [\bar{Z}_{\s,\mu}^{(1)} X_{01}(q) + Z_{\s,\mu}^{(1)} X_{10}(q)] (\delta_{\mu,6}\delta_{\s,\up}+\delta_{\mu,7}\delta_{\s,\down}) + {\rm h.c.} \notag\\
&+\frac{z_0^2}{2} [2{\rm Re}(Z_{\s,\mu}^{(1)} Z_{\s,\nu}^{(1)}) X_{11}(q) + \bar{Z}_{\s,\mu}^{(1)} Z_{\s,\nu}^{(1)} X_{20}(q) + Z_{\s,\mu}^{(1)} \bar{Z}_{\s,\nu}^{(1)} X_{02}(q)]
\Big\} \delta\psi_\nu(q) ,
\end{align}
where we introduced
\begin{align}
X_{mn}(q) = -\frac{2T}{L} \sum_k G_0(k) G_0(k+q) t_{\bf k}^m t_{\bf k+q}^n .
\end{align}

Using the fact that, at the paramagnetic saddle-point,
\begin{align}
&Z_{\s,\mu}^{(1)} = \bar{Z}_{\s,\mu}^{(1)} = Z_{\mu}^{(1)} , \quad \text{for $\mu\in\{1,2,4,5\}$}, \\
&Z_{\up,\mu}^{(1)} = Z_{\down,\mu}^{(1)} , \quad \text{for $\mu\in\{1,2,3\}$} , \\
&\bar{Z}_{\s,3}^{(1)} = -Z_{\s,3}^{(1)} , \\
&{\rm Re}(Z_{\s,\mu3}^{(2)}) = 0 , \quad \text{for $\mu\ne3$} ,
\end{align}
for all $\s\in{\up,\down}$, we finally obtain the matrix elements of the fermionic sector of the
inverse propagator
\begin{align}
D_{f,\mu\nu}^{-1}(q) &= D_{f,\nu\mu}^{-1}(q) = \frac{1}{2}z_0\xi_{\bf 0} \sum_\s Z_{\s,\mu\nu}^{(2)}
+ \frac{1}{2} \sum_\s Z_{\s,\mu}^{(1)} Z_{\s,\nu}^{(1)} \left[ \xi_{\bf q} - \frac{1}{2} z_0^2 \Pi_2(q) \right] , \quad \text{for }\mu,\nu=1,2,4,5 , \\
D_{f,\mu3}^{-1}(q) &= -D_{f,3\mu}^{-1}(q) = - \frac{{\rm i}\omega_n}{4} \sum_\s Z_{\s,\mu}^{(1)} Z_{\s,3}^{(1)} \Pi_1(q) , \quad \text{for }\mu = 1,2,4,5, \\
D_{f,33}^{-1}(q) &= z_0 \xi_{\bf 0} \left.\frac{\partial^2 {\rm Re}[z_{\s,i}(\tau)]}{\partial d''_i(\tau)\partial d''_i(\tau)}\right|_{\psi=\psi^{\rm SP}}
+ \left|\frac{\partial z_{\s,i}(\tau)}{\partial d''_i(\tau)}\right|_{\psi=\psi^{\rm SP}}^2 \left[ \xi_{\bf 0} + \frac{\omega_n^2}{2 z_0^2} \Pi_0(q) \right] , \\
D_{f,\mu6}^{-1}(q) &= D_{f,6\mu}^{-1}(q) = - \frac{1}{4} z_0 Z_{\up,\mu} \Pi_1(q) , \quad \text{for }\mu=1,2,4,5, \\
D_{f,\mu7}^{-1}(q) &= D_{f,7\mu}^{-1}(q) = - \frac{1}{4} z_0 Z_{\down,\mu} \Pi_1(q) , \quad \text{for }\mu=1,2,4,5, \\
D_{f,66}^{-1}(q) &= D_{f,77}^{-1}(q) = - \frac{1}{4} \Pi_0(q) , \\
D_{f,\mu8}^{-1}(q) &= D_{f,8\mu}^{-1}(q) = 0 , \quad \text{for all } \mu ,
\end{align}
where
\begin{align}
\Pi_m(q) = -\frac{2T}{L} \sum_k G_0(k) G_0(k+q) (t_{\bf k} + t_{\bf k+q})^m .
\end{align}

\subsubsection{Effective Gaussian theory for the slave boson fluctuations}
With the above two contributions, we arrive at the effective action functional
for the fluctuations of the slave boson fields
\begin{align}
\S_{\rm eff}[\delta\psi] = \sum_q \sum_{\mu,\nu} \delta\psi_\mu(-q) D_{\mu\nu}^{-1}(q) \delta\psi_\nu(q) , 
\end{align}
where the inverse propagator for the boson field fluctuations
$\mat{D}^{-1}(q) = \mat{D}_b^{-1}(q) + \mat{D}_f^{-1}(q)$,
is a $8\times8$ matrix in the $\delta\psi(q)$ basis.
However, if we perform a unitary change of basis
\begin{align}\label{base}
\delta\psi'(q) = \{\delta R_e(q), \delta d'(q), \delta d''(q), \delta R_0(q), \delta \beta_0(q), \delta \alpha(q), \delta R_z(q), \delta \beta_z(q)\} ,
\end{align}
where
\begin{align}
\delta R_0(q) &= \frac{1}{\sqrt{2}} [ \delta R_\up(q) + \delta R_\down(q) ] , \\
\delta R_z(q) &= \frac{1}{\sqrt{2}} [ \delta R_\up(q) - \delta R_\down(q) ] , \\
\delta \beta_0(q) &= \frac{1}{\sqrt{2}} [ \delta \beta_\up(q) + \delta \beta_\down(q) ] , \\
\delta \beta_z(q) &= \frac{1}{\sqrt{2}} [ \delta \beta_\up(q) - \delta \beta_\down(q) ] ,
\end{align}
we can decouple the inverse propagator into a pure charge fluctuations sector, with basis
\begin{align}
\delta\psi_c(q) = \{\delta R_e(q), \delta d'(q), \delta d''(q), \delta R_0(q), \delta \beta_0(q), \delta \alpha (q)\} ,
\end{align}
and a pure spin fluctuations sector, with basis
\begin{align}
\delta\psi_s(q) = \{\delta R_z(q), \delta \beta_z(q)\} .
\end{align}
To simplify notation, in the following, we refer to the fields fluctuations using
the base Eq.~(\ref{base}), and drop the prime exponent.
This means that the basis of the charge sector corresponds to $\{\delta\psi_\mu(q)\}$,
for $\mu=1,\cdots,6$, while the basis of the spin sector corresponds to $\mu=7$ and $8$.
The matrix elements for the charge sector of the fluctuation matrix then read
\begin{align}
&D^{-1}_{11}(q) = 2V_{\bf q} + D^{-1}_{f,11}(q) ,\\
&D^{-1}_{12}(q) = D^{-1}_{21}(q) = D^{-1}_{f,12}(q) ,\\
&D^{-1}_{13}(q) =-D^{-1}_{13}(q) = -\frac{{\rm i}\omega_n}{2} \Pi_1(q) Z^{(1)}_1 Z^{(1)}_3,\\
&D^{-1}_{14}(q) = D^{-1}_{41}(q) = \sqrt{2}V_{\bf q} + D^{-1}_{f,14}(q) ,\\
&D^{-1}_{15}(q) = D^{-1}_{51}(q) = -\frac{1}{2\sqrt{2}}z_0 Z^{(1)}_{1} \Pi_1(q) ,\\
&D^{-1}_{16}(q) = D^{-1}_{61}(q) = \frac{1}{2} ,\\
&D^{-1}_{22}(q) = \alpha-2\beta_0+U+D^{-1}_{f,22}(q) ,\\
&D^{-1}_{23}(q) =-D^{-1}_{32}(q) = \omega_n \left[ 1 - \frac{{\rm i} }{2} Z^{(1)}_{2} Z^{(1)}_{3} \Pi_1(q) \right] ,\\
&D^{-1}_{24}(q) = D^{-1}_{42}(q) = D^{-1}_{f,24}(q) ,\\
&D^{-1}_{25}(q) = D^{-1}_{52}(q) = -\sqrt{2}d-\frac{1}{2\sqrt{2}} z_0 Z^{(1)}_{2} \Pi_1(q) ,\\
&D^{-1}_{26}(q) = D^{-1}_{62}(q) = d ,\\
&D^{-1}_{33}(q) =  \alpha-2\beta_0+U+ z_0 Z^{(2)}_{33} \xi_{\mathbf{0}} + \left| \frac{\partial z}{\partial d''} \right|^2 \left[ \xi_{\mathbf{0}} + \frac{\omega_n^2}{2z_0^2} \Pi_0(q) \right] ,\\
&D^{-1}_{34}(q) =-D^{-1}_{43}(q) = \frac{{\rm i} \omega_n}{2} Z^{(1)}_{4} Z^{(1)}_{3} \Pi_1(q) ,\\
&D^{-1}_{35}(q) =-D^{-1}_{53}(q) = \frac{{\rm i} \omega_n}{2\sqrt{2} z_0} Z^{(1)}_{3} \Pi_0(q) ,\\
&D^{-1}_{36}(q) = D^{-1}_{63}(q) = 0 ,\\
&D^{-1}_{44}(q) = V_{\bf q} + D^{-1}_{f,4,4}(q) ,\\
&D^{-1}_{45}(q) = D^{-1}_{54}(q) = -\frac{1}{2}\left[ 1 + \frac{1}{\sqrt{2}} z_0 Z^{(1)}_{4} \Pi_1(q) \right] ,\\
&D^{-1}_{46}(q) = D^{-1}_{64}(q) = \frac{1}{\sqrt{2}} ,\\
&D^{-1}_{55}(q) = -\frac{1}{4} \Pi_0(q) ,\\
&D^{-1}_{56}(q) = D^{-1}_{65}(q) = D^{-1}_{66}(q) = 0 ,
\end{align}
where
\begin{align}
&Z^{(1)}_{\mu} = Z^{(1)}_{\up,\mu} = Z^{(1)}_{\down,\mu} , \\
&Z^{(2)}_{\mu\nu} = Z^{(2)}_{\up,\mu\nu} = Z^{(2)}_{\down,\mu\nu} , \\
&D^{-1}_{f,\mu\nu}(q) =
z_0 Z^{(2)}_{\mu\nu} \xi_{\mathbf{0}}
+ Z^{(1)}_{\mu} Z^{(1)}_{\nu} \left( \xi_{\bf q} - \frac{1}{2} z_0^2 \Pi_2(q) \right) \label{dyn:Df} .
\end{align}
While, in the spin fluctuations sector, we find
\begin{align}
&D^{-1}_{77}(q) = z_0 Z^{(2)}_{77} \xi_{\mathbf{0}} + \left( \frac{\partial z_\up}{\partial R_z} \right)^2 \left[ \xi_{\bf q} - \frac{z_0^2}{2} \Pi_2(q) \right] ,\\
&D^{-1}_{78}(q) = D^{-1}_{87}(q) = 
-\frac{1}{2} \left[ 1 + \frac{1}{\sqrt{2}} z_0 \frac{\partial z_\up}{\partial R_z} \Pi_1(q) \right] ,\\
&D^{-1}_{88}(q) = -\frac{1}{4} \Pi_0(q) ,
\end{align}
where
\begin{align}
\frac{\partial z_\up}{\partial R_z} = \left.\frac{\partial z_{\up,i}(\tau)}{\partial R_{z,i}(\tau)}\right|_{\psi=\psi^{\rm SP}} .
\end{align}
The spin sector of the inverse propagator is identical to that computed in Ref.~\cite{dao2017}
for the Hubbard model (i.e. without long range interactions), as can be checked by
expressing the derivatives of $z_\s$ with respect to the squared radial fields in terms of their
counterpart in terms of the radial fields, and using the saddle-point equations to identify
occurring $\alpha$ and $\beta_0$ terms in the obtained expressions.
In the $V\to0$ limit, the charge sector of the inverse propagator can also be shown to coincide
with that of Ref.~\cite{dao2017} by the same argument.

\subsection{Density autocorrelation function}
In this work, we are interested in the computation of the dynamical dielectric function, which
itself depends on the dynamical charge susceptibility $\chi_c({\bf q},\omega)$.
The latter may be obtained by analytical continuation of the density-density correlator
\begin{align}\label{chic1}
\chi_c(q) = \avg{n(-q)n(q)}
= 4 d^2 \avg{\delta d'(-q) \delta d'(q)} - 2 d \avg{\delta d'(-q) \delta R_{e}(q)} + \avg{\delta R_{e}(-q) \delta R_{e}(q)} .
\end{align}
Inverting the charge sector of $\mat{D}^{-1}(q)$, and using the fact that the slave boson
correlation functions are straightforwardly obtained as matrix elements of the propagator
for the slave boson fields fluctuations
\begin{align}
\avg{\delta\psi_\mu(-q)\delta\psi_\nu(q)} = \frac{1}{2}D_{\mu\nu}(q) ,
\end{align}
we obtain the explicit form of Eq.~(\ref{chic1}) as:
\begin{align}\label{dyn:chic}
\chi_c(q)=\frac{x^2 D^{-1}_{55}(q)
\left\{ s_{33} [\Delta_1(q) - 4\sqrt{2}d \Delta_2(q) + 8d^2 \Delta_3(q)] + \frac{1}{4}\omega_n^2 D^{-1}_{55}(q)\right\}}{
4s_{33} x^2 [\Delta_2^2(q) - \Delta_1(q) \Delta_3(q)] - \omega_n^2 D^{-1}_{55}(q) [\frac{1}{2} \Delta_1(q) + \sqrt{2}(x-2d)\Delta_2(q) + (x-2d)^2 \Delta_3(q)]} ,
\end{align}
where
\begin{align}
&s_{33} = \xi_{\bf 0} \left( \left|Z_3^{(1)}\right|^2 - \frac{z_0}{d} Z_2^{(1)} \right) , \\
&\Delta_{1}(q) = - D^{-1}_{55}(q)
\Big[ \frac{1}{4} D^{-1}_{22}(q) - d D^{-1}_{12}(q) + d^2 D^{-1}_{11}(q) \Big]
+ \Big[ \frac{1}{2} D^{-1}_{25}(q) - d D^{-1}_{15}(q) \Big]^2 , \\
&\Delta_{2}(q) = - D^{-1}_{55}(q)
\Big[ \frac{1}{4} D^{-1}_{24}(q) - \frac{d}{2} D^{-1}_{14}(q) - \frac{1}{2\sqrt{2}} D^{-1}_{12}(q) + \frac{d}{\sqrt{2}} D^{-1}_{11}(q) \Big] \notag\\
&\hphantom{\Delta_{2}(q)=}+ \Big[ \frac{1}{2} D^{-1}_{25}(q) - d D^{-1}_{15}(q) \Big]
\Big[ \frac{1}{2} D^{-1}_{45}(q) - \frac{1}{\sqrt{2}} D^{-1}_{15}(q) \Big] , \\
&\Delta_{3}(q) = - D^{-1}_{55}(q)
\Big[ \frac{1}{4} D^{-1}_{44}(q) - \frac{1}{\sqrt{2}} D^{-1}_{1,4}(q) + \frac{1}{2} D^{-1}_{1,1}(q) \Big]
+ \Big[ \frac{1}{2} D^{-1}_{45}(q) - \frac{1}{\sqrt{2}} D^{-1}_{15}(q) \Big]^2 .
\end{align}

\end{widetext}

\section{Slave boson mean-field and the Landau-Fermi liquid} \label{appB}
Note that, in the paramagnetic mean-field, the KRSB mapping of the
physical electron field reduces to $c_{\sigma,i} \to z_0 f_{\sigma,i}$.
Therefore, the electron's non-interacting Green's function
$G_{0,\sigma ij}(\tau)=-\avg{\mathcal{T}_\tau c_{\sigma,i}(\tau)c_{\sigma,j}^*(0)}$
may be straightforwardly expressed in terms of its similarly defined
KRSB counterpart $\tilde{G}_{0,\sigma ij}(\tau)$ as:
\begin{align}
G_{0,\sigma ij}(\tau) = z_0^2 \tilde{G}_{0,\sigma ij}(\tau) .
\end{align}
In terms of the polarizabilities $\Pi_0$ and $\Pi_0^{(0)}$, this result
translates to
\begin{align}
\Pi_0({\bf q},\omega)=\frac{1}{z_0^2} \Pi^{(0)}_0\left({\bf q},\frac{\omega}{z_0^2}\right) .
\end{align}
This equality is merely a re-statement, in terms of the KRSB inverse
effective mass renormalization factor $z_0^2$, of the standard result
from Landau's theory of the Fermi liquid that interactions renormalize
the electron gas through the quasiparticle residue~\cite{giuliani2005}.

\section{Deviation of the radial KRSB charge susceptibility from RPA}
In order to assess for the agreement, or disagreement, between the radial KRSB spectra
and the predictions of standard perturbation theory, let us compute the dynamical spin-symmetric
density interaction function $f^s({\bf q},\omega)$, defined as
\begin{align}
f^s({\bf q},\omega) = \frac{1}{\chi_c({\bf q},\omega)} - \frac{1}{\Pi_0({\bf q}, \omega)} .
\end{align}
Note that this quantity is a wavevector and frequency dependent generalization of the usual
Landau parameter $F^s_0 = N(0) f^s({\bf 0},0)$, with $N(0)$ the density of states at the Fermi
energy.
Let us also observe that $f^s$ is essentially akin to a many-body local field factor, as sometimes
introduced in early attempts to provide corrections to the RPA charge susceptibility~\cite{giuliani2005}.
In the weak coupling regime, the real part of this function should be equal to $\frac{U}{2} + V_{\bf q}$,
which would yield the standard RPA charge susceptibility, and deviation from this value thus give a
measure of the deviation of the radial KRSB results from perturbation theory.

\begin{figure}
\centering
\includegraphics[width=.47\textwidth]{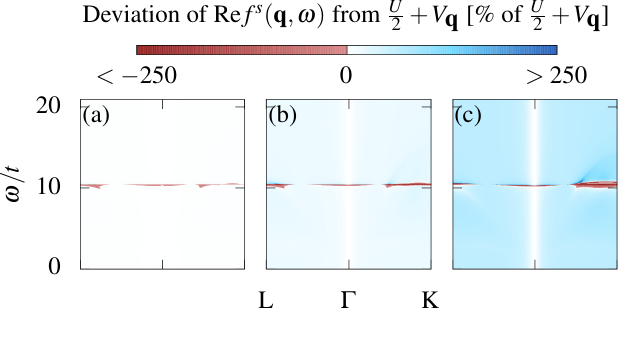}
\caption{Deviation of ${\rm Re} f^s({\bf q},\omega)$ from $\frac{U}{2} + V_{\bf q}$.
Parameters: $T=0$, and $V=0.1~U$ with (a) $U=0.01~U_c$, (b) $U=0.10~U_c$ and (c) $U=0.30~U_c$.}
\label{fig:fs}
\end{figure}

As can be seen in Fig.~\ref{fig:fs}(a), in which this deviation is displayed, for $V=0.1~U$, with
$U=0.01~U_c$, the radial KRSB results agree with standard perturbation theory in the weak coupling regime.
Indeed, we find ${\rm Re} f^s({\bf q},\omega) \simeq \frac{U}{2} + V_{\bf q}$, with deviations of at most $1.5\%$,
irrespective of the energy and wavevector, except in a narrow band around $\omega = U_c/2$.
There, deviations of up to $115\%$ are observed.
For reasons discussed in the main text, we assign this band to a signature of the upper Hubbard band (UHB),
and note that strong deviations around this UHB mode are to be expected, as it is not captured by standard perturbative
expansions.
As the Hubbard coupling is increased, we observe in Fig.~\ref{fig:fs}(b) and Fig.~\ref{fig:fs}(c) that deviations from 
${\rm Re} f^s({\bf q},\omega) \simeq \frac{U}{2} + V_{\bf q}$ grow, jointly with the deviations around
the UHB mode.\footnote{Note that, around $\Gamma$, the deviations of $f^s({\bf q},\omega)$ from
$\frac{U}{2}+V_{\bf q}$ appear to vanish. This is, however, a graphical artifact which results from
the divergence of $V_{\bf q}$ at $\Gamma$, as the data is expressed in percents of $\frac{U}{2}+V_{\bf q}$.}
In particular, we find deviations of up to $15\%$ for $U=0.10~U_c$, and $60\%$ for $U=0.30~U_c$, outside of the UHB.
Additionally the dispersion of the latter is also seen to increase, due to the larger values of $U=0.10~U_c$ and $0.30~U_c$
in Fig.~\ref{fig:fs}(b) and Fig.~\ref{fig:fs}(c), respectively.

\section{Plasma frequency on a lattice in the weak coupling regime} \label{appC}
The dispersion $\omega_{\rm plasmon}({\bf q})$ of the plasmon collective mode is obtained as a solution of
\begin{align}
\varepsilon({\bf q},\omega_{\rm plasmon}({\bf q})) = 0 .
\end{align}
In the weak coupling regime $U\ll U_c$ and $V\ll U_c$, an analytical expression for the leading contributions to
the plasmon dispersion may be obtained by noting that the radial KRSB dielectric function reduces to an RPA form
\begin{align}\label{eq-wp}
\varepsilon({\bf q},\omega) \simeq 1 + \left( \frac{U}{2} + V_{\bf q} \right) \Pi_0({\bf q},\omega) .
\end{align}
For our purpose, we expand the Lindhard function to lowest order in ${\bf q}^2$ about ${\bf q}=\Gamma$.
Re-writing it as
\begin{widetext}
\begin{align}
\Pi_0({\bf q},\omega)
= \frac{2}{L} \sum_{\bf k} \frac{n_F(E_{\bf k+q}) - n_F(E_{\bf k})}{\omega-(E_{\bf k+q}-E_{\bf k})}
= \frac{2}{L} \sum_{\bf k} n_F(E_{\bf k})
\frac{2 E_{\bf k} - E_{\bf k+q} - E_{\bf k-q}}{\omega^2 + \omega ( E_{\bf k-q} - E_{\bf k+q} ) + (E_{\bf k+q} - E_{\bf k}) (E_{\bf k} - E_{\bf k-q})} ,
\end{align}
\end{widetext}
then expanding the dispersion as 
$E_{\bf k \pm q} =  E_{\bf k} \pm {\bf q}^{\rm T}\nabla_{\bf k}E_{\bf k} + \frac{1}{2} {\bf q}^{\rm T}{\bf H}(E_{\bf k}){\bf q} + O(q^3)$, with
${\bf H}(E_{\bf k})$ the Hessian matrix $H_{ab}(E_{\bf k})=\partial^2E_{\bf k}/\partial k_a \partial k_b$,
and using $\omega \gg |{\bf q}| \equiv q$ (since $\omega_p$ remains finite for finite $V$), we find
\begin{align}
\Pi_0({\bf q},\omega) \simeq - \frac{2}{L} \sum_{\bf k} n_F(E_{\bf k})
\frac{{\bf q}^{\rm T}{\bf H}(E_{\bf k}){\bf q}}{\omega^2} .
\end{align}
We can moreover make use of the fact that the off-diagonal matrix elements of the Hessian are odd with respect
to the components of ${\bf k}$, as well as  the invariance of the remaining integrals under permutations
of the indices of ${\bf H}$, such that we end up with the simple form
\begin{align}
\Pi_0({\bf q},\omega)
&= -\frac{2}{L} \sum_{\bf k} n_F(E_{\bf k}) \frac{\sum_a q_a^2 H_{aa}(E_{\bf k})}{\omega^2} \notag\\
&= -\frac{q^2}{\omega^2} \frac{2t}{L} \sum_{\bf k} n_F(E_{\bf k}) \cos\frac{k_x}{2} \! \left( \! \cos\frac{k_y}{2} \! + \! \cos\frac{k_z}{2} \! \right) \notag\\
&= \frac{q^2}{\omega^2} \frac{\xi_{\bf 0}}{6} ,
\end{align}
where $\xi_{\bf 0}$ is the bare average kinetic energy per lattice site.
Inserting this expression into Eq.~(\ref{eq-wp}), we finally find
\begin{align}
\omega_{\rm plasmon}^2({\bf q}) \simeq \omega_p^2 + \kappa q^2 ,
\end{align}
with
\begin{align}
\omega_p^2 = -\frac{V \xi_{\bf 0}}{6} ,
\end{align}
and
\begin{align}
\kappa = -\frac{U \xi_{\bf 0}}{12} .
\end{align}
$V$ is hence pivotal to the very existence of the plasmon altogether, while $U$ rather governs its dispersion.
Moreover, both $\omega_p$ and $\kappa$ are sensitive to the lattice on which the electrons evolve through the
$-\xi_{\bf 0}/6$ factor.
Finally, recalling that in free space $V=e^2$, and that the kinetic energy
is proportional to the inverse band mass via $t\sim1/m^*$, we may thus rewrite 
$\xi_{\bf 0} \sim -6~\varrho/m^*$, with
\begin{align}
\varrho
&\equiv \frac{2}{L t} \sum_{\bf k} n_F(E_{\bf k}) \frac{\partial^2 E_{\bf k}}{\partial k_x^2} \notag\\
&= \frac{2}{L} \sum_{\bf k} n_F(E_{\bf k}) \cos\frac{k_x}{2} \left( \cos\frac{k_y}{2} + \cos\frac{k_z}{2} \right) .
\end{align}
The plasma frequency is then recast as
\begin{align}
\omega_p = \sqrt{\frac{4\pi e^2 \varrho}{m^*}} .
\end{align}
This is the classical expression, apart from the fact that the electron density $n$
has been replaced by $\varrho$.
The difference stems from the way we represent the density distribution on the lattice.
In contrast to the Fermi gas, for which a homogeneous electron density is given by a continuous (constant
in this case) function of the position, we here deal with a discretized and periodic function of the position
${\bf R}$:
$n({\bf R}) \sim \sum_{j} \delta({\bf R}-{\bf r}_j)$, where ${\bf r}_j$ is a lattice vector.
The Coulomb potential then couples to this set of discrete and periodic lattice bonds, and
taking its Fourier transform results in contributions from the lattice harmonics.
Therefore, the lattice-dependent $\varrho$ appears in the plasma frequency instead of
$n=2 \int \frac{{\rm d}^3 \bf p}{(2\pi)^3} n_F(\frac{p^2}{2m^*})$ for the Fermi gas.


\begin{thebibliography}{71}%
\makeatletter
\providecommand \@ifxundefined [1]{%
 \@ifx{#1\undefined}
}%
\providecommand \@ifnum [1]{%
 \ifnum #1\expandafter \@firstoftwo
 \else \expandafter \@secondoftwo
 \fi
}%
\providecommand \@ifx [1]{%
 \ifx #1\expandafter \@firstoftwo
 \else \expandafter \@secondoftwo
 \fi
}%
\providecommand \natexlab [1]{#1}%
\providecommand \enquote  [1]{``#1''}%
\providecommand \bibnamefont  [1]{#1}%
\providecommand \bibfnamefont [1]{#1}%
\providecommand \citenamefont [1]{#1}%
\providecommand \href@noop [0]{\@secondoftwo}%
\providecommand \href [0]{\begingroup \@sanitize@url \@href}%
\providecommand \@href[1]{\@@startlink{#1}\@@href}%
\providecommand \@@href[1]{\endgroup#1\@@endlink}%
\providecommand \@sanitize@url [0]{\catcode `\\12\catcode `\$12\catcode
  `\&12\catcode `\#12\catcode `\^12\catcode `\_12\catcode `\%12\relax}%
\providecommand \@@startlink[1]{}%
\providecommand \@@endlink[0]{}%
\providecommand \url  [0]{\begingroup\@sanitize@url \@url }%
\providecommand \@url [1]{\endgroup\@href {#1}{\urlprefix }}%
\providecommand \urlprefix  [0]{URL }%
\providecommand \Eprint [0]{\href }%
\providecommand \doibase [0]{https://doi.org/}%
\providecommand \selectlanguage [0]{\@gobble}%
\providecommand \bibinfo  [0]{\@secondoftwo}%
\providecommand \bibfield  [0]{\@secondoftwo}%
\providecommand \translation [1]{[#1]}%
\providecommand \BibitemOpen [0]{}%
\providecommand \bibitemStop [0]{}%
\providecommand \bibitemNoStop [0]{.\EOS\space}%
\providecommand \EOS [0]{\spacefactor3000\relax}%
\providecommand \BibitemShut  [1]{\csname bibitem#1\endcsname}%
\let\auto@bib@innerbib\@empty
\bibitem [{\citenamefont {Bohm}\ and\ \citenamefont {Pines}(1951)}]{bohm1951}%
  \BibitemOpen
  \bibfield  {author} {\bibinfo {author} {\bibfnamefont {D.}~\bibnamefont
  {Bohm}}\ and\ \bibinfo {author} {\bibfnamefont {D.}~\bibnamefont {Pines}},\
  }\bibfield  {title} {\bibinfo {title} {A collective description of electron
  interactions. I. magnetic interactions},\ }\href
  {https://doi.org/10.1103/PhysRev.82.625} {\bibfield  {journal} {\bibinfo
  {journal} {Phys. Rev.}\ }\textbf {\bibinfo {volume} {82}},\ \bibinfo {pages}
  {625} (\bibinfo {year} {1951})}\BibitemShut {NoStop}%
\bibitem [{\citenamefont {Pines}\ and\ \citenamefont {Bohm}(1952)}]{pines1951}%
  \BibitemOpen
  \bibfield  {author} {\bibinfo {author} {\bibfnamefont {D.}~\bibnamefont
  {Pines}}\ and\ \bibinfo {author} {\bibfnamefont {D.}~\bibnamefont {Bohm}},\
  }\bibfield  {title} {\bibinfo {title} {A collective description of electron
  interactions: {II}. collective $\mathrm{vs}$ individual particle aspects of
  the interactions},\ }\href {https://doi.org/10.1103/PhysRev.85.338}
  {\bibfield  {journal} {\bibinfo  {journal} {Phys. Rev.}\ }\textbf {\bibinfo
  {volume} {85}},\ \bibinfo {pages} {338} (\bibinfo {year} {1952})}\BibitemShut
  {NoStop}%
\bibitem [{\citenamefont {Bohm}\ and\ \citenamefont {Pines}(1953)}]{bohm1953}%
  \BibitemOpen
  \bibfield  {author} {\bibinfo {author} {\bibfnamefont {D.}~\bibnamefont
  {Bohm}}\ and\ \bibinfo {author} {\bibfnamefont {D.}~\bibnamefont {Pines}},\
  }\bibfield  {title} {\bibinfo {title} {A collective description of electron
  interactions: III. coulomb interactions in a degenerate electron gas},\
  }\href {https://doi.org/10.1103/PhysRev.92.609} {\bibfield  {journal}
  {\bibinfo  {journal} {Phys. Rev.}\ }\textbf {\bibinfo {volume} {92}},\
  \bibinfo {pages} {609} (\bibinfo {year} {1953})}\BibitemShut {NoStop}%
\bibitem [{\citenamefont {Anderson}(1994)}]{anderson1994}%
  \BibitemOpen
  \bibfield  {author} {\bibinfo {author} {\bibfnamefont {P.~W.}\ \bibnamefont
  {Anderson}},\ }\href {https://doi.org/10.4324/9780429494116} {\emph {\bibinfo
  {title} {Basic Notions of Condensed Matter Physics}}}\ (\bibinfo  {publisher}
  {Addison-Wesley, Redwood City, CA},\ \bibinfo {year} {1994})\BibitemShut
  {NoStop}%
\bibitem [{\citenamefont {Schrieffer}(1964)}]{schrieffer1999}%
  \BibitemOpen
  \bibfield  {author} {\bibinfo {author} {\bibfnamefont {J.~R.}\ \bibnamefont
  {Schrieffer}},\ }\href {https://doi.org/10.1201/9780429495700} {\emph
  {\bibinfo {title} {Theory of Superconductivity}}}\ (\bibinfo  {publisher}
  {Frontiers in Physics, Addison-Wesley, New York, NY},\ \bibinfo {year}
  {1964})\BibitemShut {NoStop}%
\bibitem [{\citenamefont {Giuliani}\ and\ \citenamefont
  {Vignale}(2005)}]{giuliani2005}%
  \BibitemOpen
  \bibfield  {author} {\bibinfo {author} {\bibfnamefont {G.}~\bibnamefont
  {Giuliani}}\ and\ \bibinfo {author} {\bibfnamefont {G.}~\bibnamefont
  {Vignale}},\ }\href {https://doi.org/10.1017/CBO9780511619915} {\emph
  {\bibinfo {title} {Quantum Theory of the Electron Liquid}}}\ (\bibinfo
  {publisher} {Cambridge University Press},\ \bibinfo {year}
  {2005})\BibitemShut {NoStop}%
\bibitem [{\citenamefont {Yan}\ and\ \citenamefont
  {Mortensen}(2016)}]{yan2016}%
  \BibitemOpen
  \bibfield  {author} {\bibinfo {author} {\bibfnamefont {W.}~\bibnamefont
  {Yan}}\ and\ \bibinfo {author} {\bibfnamefont {N.~A.}\ \bibnamefont
  {Mortensen}},\ }\bibfield  {title} {\bibinfo {title} {Nonclassical effects in
  plasmonics: An energy perspective to quantify nonclassical effects},\ }\href
  {https://doi.org/10.1103/PhysRevB.93.115439} {\bibfield  {journal} {\bibinfo
  {journal} {Phys. Rev. B}\ }\textbf {\bibinfo {volume} {93}},\ \bibinfo
  {pages} {115439} (\bibinfo {year} {2016})}\BibitemShut {NoStop}%
\bibitem [{\citenamefont {Chang}\ \emph {et~al.}(2007)\citenamefont {Chang},
  \citenamefont {Sørensen}, \citenamefont {Demler},\ and\ \citenamefont
  {Lukin}}]{chang2007}%
  \BibitemOpen
  \bibfield  {author} {\bibinfo {author} {\bibfnamefont {D.~E.}\ \bibnamefont
  {Chang}}, \bibinfo {author} {\bibfnamefont {A.~S.}\ \bibnamefont
  {Sørensen}}, \bibinfo {author} {\bibfnamefont {E.~A.}\ \bibnamefont
  {Demler}},\ and\ \bibinfo {author} {\bibfnamefont {M.~D.}\ \bibnamefont
  {Lukin}},\ }\bibfield  {title} {\bibinfo {title} {A single-photon transistor
  using nanoscale surface plasmons},\ }\href {https://doi.org/10.1038/nphys708}
  {\bibfield  {journal} {\bibinfo  {journal} {Nat. Phys.}\ }\textbf {\bibinfo
  {volume} {3}},\ \bibinfo {pages} {807} (\bibinfo {year} {2007})}\BibitemShut
  {NoStop}%
\bibitem [{\citenamefont {Lezec}\ \emph {et~al.}(2007)\citenamefont {Lezec},
  \citenamefont {Dionne},\ and\ \citenamefont {Atwater}}]{lezec2007}%
  \BibitemOpen
  \bibfield  {author} {\bibinfo {author} {\bibfnamefont {H.~J.}\ \bibnamefont
  {Lezec}}, \bibinfo {author} {\bibfnamefont {J.~A.}\ \bibnamefont {Dionne}},\
  and\ \bibinfo {author} {\bibfnamefont {H.~A.}\ \bibnamefont {Atwater}},\
  }\bibfield  {title} {\bibinfo {title} {Negative refraction at visible
  frequencies},\ }\href {https://doi.org/10.1126/science.1139266} {\bibfield
  {journal} {\bibinfo  {journal} {Science}\ }\textbf {\bibinfo {volume}
  {316}},\ \bibinfo {pages} {430} (\bibinfo {year} {2007})}\BibitemShut
  {NoStop}%
\bibitem [{\citenamefont {Akimov}\ \emph {et~al.}(2007)\citenamefont {Akimov},
  \citenamefont {Mukherjee}, \citenamefont {Yu}, \citenamefont {Chang},
  \citenamefont {Zibrov}, \citenamefont {Hemmer}, \citenamefont {Park},\ and\
  \citenamefont {Lukin}}]{akimov2007}%
  \BibitemOpen
  \bibfield  {author} {\bibinfo {author} {\bibfnamefont {A.~V.}\ \bibnamefont
  {Akimov}}, \bibinfo {author} {\bibfnamefont {A.}~\bibnamefont {Mukherjee}},
  \bibinfo {author} {\bibfnamefont {C.~L.}\ \bibnamefont {Yu}}, \bibinfo
  {author} {\bibfnamefont {D.~E.}\ \bibnamefont {Chang}}, \bibinfo {author}
  {\bibfnamefont {A.~S.}\ \bibnamefont {Zibrov}}, \bibinfo {author}
  {\bibfnamefont {P.~R.}\ \bibnamefont {Hemmer}}, \bibinfo {author}
  {\bibfnamefont {H.}~\bibnamefont {Park}},\ and\ \bibinfo {author}
  {\bibfnamefont {M.~D.}\ \bibnamefont {Lukin}},\ }\bibfield  {title} {\bibinfo
  {title} {Generation of single optical plasmons in metallic nanowires coupled
  to quantum dots},\ }\href {https://doi.org/10.1038/nature06230} {\bibfield
  {journal} {\bibinfo  {journal} {Nature}\ }\textbf {\bibinfo {volume} {450}},\
  \bibinfo {pages} {402} (\bibinfo {year} {2007})}\BibitemShut {NoStop}%
\bibitem [{\citenamefont {Noginov}\ \emph {et~al.}(2009)\citenamefont
  {Noginov}, \citenamefont {Zhu}, \citenamefont {Belgrave}, \citenamefont
  {Bakker}, \citenamefont {Shalaev}, \citenamefont {Narimanov}, \citenamefont
  {Stout}, \citenamefont {Herz}, \citenamefont {Suteewong},\ and\ \citenamefont
  {Wiesner}}]{noginov2009}%
  \BibitemOpen
  \bibfield  {author} {\bibinfo {author} {\bibfnamefont {M.~A.}\ \bibnamefont
  {Noginov}}, \bibinfo {author} {\bibfnamefont {G.}~\bibnamefont {Zhu}},
  \bibinfo {author} {\bibfnamefont {A.~M.}\ \bibnamefont {Belgrave}}, \bibinfo
  {author} {\bibfnamefont {R.}~\bibnamefont {Bakker}}, \bibinfo {author}
  {\bibfnamefont {V.~M.}\ \bibnamefont {Shalaev}}, \bibinfo {author}
  {\bibfnamefont {E.~E.}\ \bibnamefont {Narimanov}}, \bibinfo {author}
  {\bibfnamefont {S.}~\bibnamefont {Stout}}, \bibinfo {author} {\bibfnamefont
  {E.}~\bibnamefont {Herz}}, \bibinfo {author} {\bibfnamefont {T.}~\bibnamefont
  {Suteewong}},\ and\ \bibinfo {author} {\bibfnamefont {U.}~\bibnamefont
  {Wiesner}},\ }\bibfield  {title} {\bibinfo {title} {Demonstration of a
  spaser-based nanolaser},\ }\href {https://doi.org/10.1038/nature08318}
  {\bibfield  {journal} {\bibinfo  {journal} {Nature}\ }\textbf {\bibinfo
  {volume} {460}},\ \bibinfo {pages} {1110} (\bibinfo {year}
  {2009})}\BibitemShut {NoStop}%
\bibitem [{\citenamefont {Oulton}\ \emph {et~al.}(2009)\citenamefont {Oulton},
  \citenamefont {Sorger}, \citenamefont {Zentgraf}, \citenamefont {Ma},
  \citenamefont {Gladden}, \citenamefont {Dai}, \citenamefont {Bartal},\ and\
  \citenamefont {Zhang}}]{oulton2009}%
  \BibitemOpen
  \bibfield  {author} {\bibinfo {author} {\bibfnamefont {R.~F.}\ \bibnamefont
  {Oulton}}, \bibinfo {author} {\bibfnamefont {V.~J.}\ \bibnamefont {Sorger}},
  \bibinfo {author} {\bibfnamefont {T.}~\bibnamefont {Zentgraf}}, \bibinfo
  {author} {\bibfnamefont {R.-M.}\ \bibnamefont {Ma}}, \bibinfo {author}
  {\bibfnamefont {C.}~\bibnamefont {Gladden}}, \bibinfo {author} {\bibfnamefont
  {L.}~\bibnamefont {Dai}}, \bibinfo {author} {\bibfnamefont {G.}~\bibnamefont
  {Bartal}},\ and\ \bibinfo {author} {\bibfnamefont {X.}~\bibnamefont
  {Zhang}},\ }\bibfield  {title} {\bibinfo {title} {Plasmon lasers at deep
  subwavelength scale},\ }\href {https://doi.org/10.1038/nature08364}
  {\bibfield  {journal} {\bibinfo  {journal} {Nature}\ }\textbf {\bibinfo
  {volume} {461}},\ \bibinfo {pages} {629} (\bibinfo {year}
  {2009})}\BibitemShut {NoStop}%
\bibitem [{\citenamefont {Cai}\ \emph {et~al.}(2009)\citenamefont {Cai},
  \citenamefont {White},\ and\ \citenamefont {Brongersma}}]{cai2009}%
  \BibitemOpen
  \bibfield  {author} {\bibinfo {author} {\bibfnamefont {W.}~\bibnamefont
  {Cai}}, \bibinfo {author} {\bibfnamefont {J.~S.}\ \bibnamefont {White}},\
  and\ \bibinfo {author} {\bibfnamefont {M.~L.}\ \bibnamefont {Brongersma}},\
  }\bibfield  {title} {\bibinfo {title} {Compact, high-speed and
  power-efficient electrooptic plasmonic modulators},\ }\href
  {https://doi.org/10.1021/nl902701b} {\bibfield  {journal} {\bibinfo
  {journal} {Nano Lett.}\ }\textbf {\bibinfo {volume} {9}},\ \bibinfo {pages}
  {4403} (\bibinfo {year} {2009})}\BibitemShut {NoStop}%
\bibitem [{\citenamefont {Hryciw}\ \emph {et~al.}(2010)\citenamefont {Hryciw},
  \citenamefont {Jun},\ and\ \citenamefont {Brongersma}}]{hryciw2010}%
  \BibitemOpen
  \bibfield  {author} {\bibinfo {author} {\bibfnamefont {A.}~\bibnamefont
  {Hryciw}}, \bibinfo {author} {\bibfnamefont {Y.~C.}\ \bibnamefont {Jun}},\
  and\ \bibinfo {author} {\bibfnamefont {M.~L.}\ \bibnamefont {Brongersma}},\
  }\bibfield  {title} {\bibinfo {title} {Electrifying plasmonics on silicon},\
  }\href {https://doi.org/10.1038/nmat2598} {\bibfield  {journal} {\bibinfo
  {journal} {Nat. Mater.}\ }\textbf {\bibinfo {volume} {9}},\ \bibinfo {pages}
  {3} (\bibinfo {year} {2010})}\BibitemShut {NoStop}%
\bibitem [{\citenamefont {Schuller}\ \emph {et~al.}(2010)\citenamefont
  {Schuller}, \citenamefont {Barnard}, \citenamefont {Cai}, \citenamefont
  {Jun}, \citenamefont {White},\ and\ \citenamefont
  {Brongersma}}]{schuller2010}%
  \BibitemOpen
  \bibfield  {author} {\bibinfo {author} {\bibfnamefont {J.~A.}\ \bibnamefont
  {Schuller}}, \bibinfo {author} {\bibfnamefont {E.~S.}\ \bibnamefont
  {Barnard}}, \bibinfo {author} {\bibfnamefont {W.}~\bibnamefont {Cai}},
  \bibinfo {author} {\bibfnamefont {Y.~C.}\ \bibnamefont {Jun}}, \bibinfo
  {author} {\bibfnamefont {J.~S.}\ \bibnamefont {White}},\ and\ \bibinfo
  {author} {\bibfnamefont {M.~L.}\ \bibnamefont {Brongersma}},\ }\bibfield
  {title} {\bibinfo {title} {Plasmonics for extreme light concentration and
  manipulation},\ }\href {https://doi.org/10.1038/nmat2630} {\bibfield
  {journal} {\bibinfo  {journal} {Nat. Mater.}\ }\textbf {\bibinfo {volume}
  {9}},\ \bibinfo {pages} {193} (\bibinfo {year} {2010})}\BibitemShut {NoStop}%
\bibitem [{\citenamefont {Ergin}\ \emph {et~al.}(2010)\citenamefont {Ergin},
  \citenamefont {Stenger}, \citenamefont {Brenner}, \citenamefont {Pendry},\
  and\ \citenamefont {Wegener}}]{ergin2010}%
  \BibitemOpen
  \bibfield  {author} {\bibinfo {author} {\bibfnamefont {T.}~\bibnamefont
  {Ergin}}, \bibinfo {author} {\bibfnamefont {N.}~\bibnamefont {Stenger}},
  \bibinfo {author} {\bibfnamefont {P.}~\bibnamefont {Brenner}}, \bibinfo
  {author} {\bibfnamefont {J.~B.}\ \bibnamefont {Pendry}},\ and\ \bibinfo
  {author} {\bibfnamefont {M.}~\bibnamefont {Wegener}},\ }\bibfield  {title}
  {\bibinfo {title} {Three-dimensional invisibility cloak at optical
  wavelengths},\ }\href {https://doi.org/10.1126/science.1186351} {\bibfield
  {journal} {\bibinfo  {journal} {Science}\ }\textbf {\bibinfo {volume}
  {328}},\ \bibinfo {pages} {337} (\bibinfo {year} {2010})}\BibitemShut
  {NoStop}%
\bibitem [{\citenamefont {Mayer}\ and\ \citenamefont
  {Hafner}(2011)}]{mayer2011}%
  \BibitemOpen
  \bibfield  {author} {\bibinfo {author} {\bibfnamefont {K.~M.}\ \bibnamefont
  {Mayer}}\ and\ \bibinfo {author} {\bibfnamefont {J.~H.}\ \bibnamefont
  {Hafner}},\ }\bibfield  {title} {\bibinfo {title} {Localized surface plasmon
  resonance sensors},\ }\href {https://doi.org/10.1021/cr100313v} {\bibfield
  {journal} {\bibinfo  {journal} {Chem. Rev.}\ }\textbf {\bibinfo {volume}
  {111}},\ \bibinfo {pages} {3828} (\bibinfo {year} {2011})}\BibitemShut
  {NoStop}%
\bibitem [{\citenamefont {Boltasseva}\ and\ \citenamefont
  {Atwater}(2011)}]{botlasseva2011}%
  \BibitemOpen
  \bibfield  {author} {\bibinfo {author} {\bibfnamefont {A.}~\bibnamefont
  {Boltasseva}}\ and\ \bibinfo {author} {\bibfnamefont {H.~A.}\ \bibnamefont
  {Atwater}},\ }\bibfield  {title} {\bibinfo {title} {Low-loss plasmonic
  metamaterials},\ }\href {https://doi.org/10.1126/science.1198258} {\bibfield
  {journal} {\bibinfo  {journal} {Science}\ }\textbf {\bibinfo {volume}
  {331}},\ \bibinfo {pages} {290} (\bibinfo {year} {2011})}\BibitemShut
  {NoStop}%
\bibitem [{\citenamefont {Soukoulis}\ and\ \citenamefont
  {Wegener}(2011)}]{soukoulis2011}%
  \BibitemOpen
  \bibfield  {author} {\bibinfo {author} {\bibfnamefont {C.~M.}\ \bibnamefont
  {Soukoulis}}\ and\ \bibinfo {author} {\bibfnamefont {M.}~\bibnamefont
  {Wegener}},\ }\bibfield  {title} {\bibinfo {title} {Past achievements and
  future challenges in the development of three-dimensional photonic
  metamaterials},\ }\href {https://doi.org/10.1038/nphoton.2011.154} {\bibfield
   {journal} {\bibinfo  {journal} {Nat. Photon.}\ }\textbf {\bibinfo {volume}
  {5}},\ \bibinfo {pages} {523} (\bibinfo {year} {2011})}\BibitemShut {NoStop}%
\bibitem [{\citenamefont {Huang}\ \emph {et~al.}(2014)\citenamefont {Huang},
  \citenamefont {Seo}, \citenamefont {Sarmiento}, \citenamefont {Huo},
  \citenamefont {Harris},\ and\ \citenamefont {Brongersma}}]{huang2014}%
  \BibitemOpen
  \bibfield  {author} {\bibinfo {author} {\bibfnamefont {K.~C.~Y.}\
  \bibnamefont {Huang}}, \bibinfo {author} {\bibfnamefont {M.-K.}\ \bibnamefont
  {Seo}}, \bibinfo {author} {\bibfnamefont {T.}~\bibnamefont {Sarmiento}},
  \bibinfo {author} {\bibfnamefont {Y.}~\bibnamefont {Huo}}, \bibinfo {author}
  {\bibfnamefont {J.~S.}\ \bibnamefont {Harris}},\ and\ \bibinfo {author}
  {\bibfnamefont {M.~L.}\ \bibnamefont {Brongersma}},\ }\bibfield  {title}
  {\bibinfo {title} {Electrically driven subwavelength optical nanocircuits},\
  }\href {https://doi.org/10.1038/nphoton.2014.2} {\bibfield  {journal}
  {\bibinfo  {journal} {Nat. Photon.}\ }\textbf {\bibinfo {volume} {8}},\
  \bibinfo {pages} {244} (\bibinfo {year} {2014})}\BibitemShut {NoStop}%
\bibitem [{\citenamefont {Baev}\ \emph {et~al.}(2015)\citenamefont {Baev},
  \citenamefont {Prasad}, \citenamefont {Ågren}, \citenamefont {Samoć},\ and\
  \citenamefont {Wegener}}]{baev2015}%
  \BibitemOpen
  \bibfield  {author} {\bibinfo {author} {\bibfnamefont {A.}~\bibnamefont
  {Baev}}, \bibinfo {author} {\bibfnamefont {P.~N.}\ \bibnamefont {Prasad}},
  \bibinfo {author} {\bibfnamefont {H.}~\bibnamefont {Ågren}}, \bibinfo
  {author} {\bibfnamefont {M.}~\bibnamefont {Samoć}},\ and\ \bibinfo {author}
  {\bibfnamefont {M.}~\bibnamefont {Wegener}},\ }\bibfield  {title} {\bibinfo
  {title} {Metaphotonics: An emerging field with opportunities and
  challenges},\ }\href
  {https://doi.org/https://doi.org/10.1016/j.physrep.2015.07.002} {\bibfield
  {journal} {\bibinfo  {journal} {Phys. Rep.}\ }\textbf {\bibinfo {volume}
  {594}},\ \bibinfo {pages} {1} (\bibinfo {year} {2015})}\BibitemShut {NoStop}%
\bibitem [{\citenamefont {Atwater}\ and\ \citenamefont
  {Polman}(2010)}]{atwater2010}%
  \BibitemOpen
  \bibfield  {author} {\bibinfo {author} {\bibfnamefont {H.~A.}\ \bibnamefont
  {Atwater}}\ and\ \bibinfo {author} {\bibfnamefont {A.}~\bibnamefont
  {Polman}},\ }\bibfield  {title} {\bibinfo {title} {Plasmonics for improved
  photovoltaic devices},\ }\href {https://doi.org/10.1038/nmat2629} {\bibfield
  {journal} {\bibinfo  {journal} {Nat. Mater.}\ }\textbf {\bibinfo {volume}
  {9}},\ \bibinfo {pages} {205} (\bibinfo {year} {2010})}\BibitemShut {NoStop}%
\bibitem [{\citenamefont {Linic}\ \emph {et~al.}(2011)\citenamefont {Linic},
  \citenamefont {Christopher},\ and\ \citenamefont {Ingram}}]{linic2011}%
  \BibitemOpen
  \bibfield  {author} {\bibinfo {author} {\bibfnamefont {S.}~\bibnamefont
  {Linic}}, \bibinfo {author} {\bibfnamefont {P.}~\bibnamefont {Christopher}},\
  and\ \bibinfo {author} {\bibfnamefont {D.~B.}\ \bibnamefont {Ingram}},\
  }\bibfield  {title} {\bibinfo {title} {Plasmonic-metal nanostructures for
  efficient conversion of solar to chemical energy},\ }\href
  {https://doi.org/10.1038/nmat3151} {\bibfield  {journal} {\bibinfo  {journal}
  {Nat. Mater.}\ }\textbf {\bibinfo {volume} {10}},\ \bibinfo {pages} {911}
  (\bibinfo {year} {2011})}\BibitemShut {NoStop}%
\bibitem [{\citenamefont {Hou}\ \emph {et~al.}(2011)\citenamefont {Hou},
  \citenamefont {Hung}, \citenamefont {Pavaskar}, \citenamefont {Goeppert},
  \citenamefont {Aykol},\ and\ \citenamefont {Cronin}}]{hou2011}%
  \BibitemOpen
  \bibfield  {author} {\bibinfo {author} {\bibfnamefont {W.}~\bibnamefont
  {Hou}}, \bibinfo {author} {\bibfnamefont {W.~H.}\ \bibnamefont {Hung}},
  \bibinfo {author} {\bibfnamefont {P.}~\bibnamefont {Pavaskar}}, \bibinfo
  {author} {\bibfnamefont {A.}~\bibnamefont {Goeppert}}, \bibinfo {author}
  {\bibfnamefont {M.}~\bibnamefont {Aykol}},\ and\ \bibinfo {author}
  {\bibfnamefont {S.~B.}\ \bibnamefont {Cronin}},\ }\bibfield  {title}
  {\bibinfo {title} {Photocatalytic conversion of {CO}$_2$ to hydrocarbon fuels
  via plasmon-enhanced absorption and metallic interband transitions},\ }\href
  {https://doi.org/10.1021/cs2001434} {\bibfield  {journal} {\bibinfo
  {journal} {ACS Catal.}\ }\textbf {\bibinfo {volume} {1}},\ \bibinfo {pages}
  {929} (\bibinfo {year} {2011})}\BibitemShut {NoStop}%
\bibitem [{\citenamefont {Lee}\ \emph {et~al.}(2012)\citenamefont {Lee},
  \citenamefont {Mubeen}, \citenamefont {Ji}, \citenamefont {Stucky},\ and\
  \citenamefont {Moskovits}}]{lee2012}%
  \BibitemOpen
  \bibfield  {author} {\bibinfo {author} {\bibfnamefont {J.}~\bibnamefont
  {Lee}}, \bibinfo {author} {\bibfnamefont {S.}~\bibnamefont {Mubeen}},
  \bibinfo {author} {\bibfnamefont {X.}~\bibnamefont {Ji}}, \bibinfo {author}
  {\bibfnamefont {G.~D.}\ \bibnamefont {Stucky}},\ and\ \bibinfo {author}
  {\bibfnamefont {M.}~\bibnamefont {Moskovits}},\ }\bibfield  {title} {\bibinfo
  {title} {Plasmonic photoanodes for solar water splitting with visible
  light},\ }\href {https://doi.org/10.1021/nl302796f} {\bibfield  {journal}
  {\bibinfo  {journal} {Nano Lett.}\ }\textbf {\bibinfo {volume} {12}},\
  \bibinfo {pages} {5014} (\bibinfo {year} {2012})}\BibitemShut {NoStop}%
\bibitem [{\citenamefont {Mukherjee}\ \emph {et~al.}(2013)\citenamefont
  {Mukherjee}, \citenamefont {Libisch}, \citenamefont {Large}, \citenamefont
  {Neumann}, \citenamefont {Brown}, \citenamefont {Cheng}, \citenamefont
  {Lassiter}, \citenamefont {Carter}, \citenamefont {Nordlander},\ and\
  \citenamefont {Halas}}]{mukherjee2013}%
  \BibitemOpen
  \bibfield  {author} {\bibinfo {author} {\bibfnamefont {S.}~\bibnamefont
  {Mukherjee}}, \bibinfo {author} {\bibfnamefont {F.}~\bibnamefont {Libisch}},
  \bibinfo {author} {\bibfnamefont {N.}~\bibnamefont {Large}}, \bibinfo
  {author} {\bibfnamefont {O.}~\bibnamefont {Neumann}}, \bibinfo {author}
  {\bibfnamefont {L.~V.}\ \bibnamefont {Brown}}, \bibinfo {author}
  {\bibfnamefont {J.}~\bibnamefont {Cheng}}, \bibinfo {author} {\bibfnamefont
  {J.~B.}\ \bibnamefont {Lassiter}}, \bibinfo {author} {\bibfnamefont {E.~A.}\
  \bibnamefont {Carter}}, \bibinfo {author} {\bibfnamefont {P.}~\bibnamefont
  {Nordlander}},\ and\ \bibinfo {author} {\bibfnamefont {N.~J.}\ \bibnamefont
  {Halas}},\ }\bibfield  {title} {\bibinfo {title} {Hot electrons do the
  impossible: Plasmon-induced dissociation of {H}$_2$ on {Au}},\ }\href
  {https://doi.org/10.1021/nl303940z} {\bibfield  {journal} {\bibinfo
  {journal} {Nano Lett.}\ }\textbf {\bibinfo {volume} {13}},\ \bibinfo {pages}
  {240} (\bibinfo {year} {2013})}\BibitemShut {NoStop}%
\bibitem [{\citenamefont {O'Neal}\ \emph {et~al.}(2004)\citenamefont {O'Neal},
  \citenamefont {Hirsch}, \citenamefont {Halas}, \citenamefont {Payne},\ and\
  \citenamefont {West}}]{oneal2004}%
  \BibitemOpen
  \bibfield  {author} {\bibinfo {author} {\bibfnamefont {D.}~\bibnamefont
  {O'Neal}}, \bibinfo {author} {\bibfnamefont {L.~R.}\ \bibnamefont {Hirsch}},
  \bibinfo {author} {\bibfnamefont {N.~J.}\ \bibnamefont {Halas}}, \bibinfo
  {author} {\bibfnamefont {J.}~\bibnamefont {Payne}},\ and\ \bibinfo {author}
  {\bibfnamefont {J.~L.}\ \bibnamefont {West}},\ }\bibfield  {title} {\bibinfo
  {title} {Photo-thermal tumor ablation in mice using near infrared-absorbing
  nanoparticles},\ }\href {https://doi.org/10.1016/j.canlet.2004.02.004}
  {\bibfield  {journal} {\bibinfo  {journal} {Cancer Lett.}\ }\textbf {\bibinfo
  {volume} {209}},\ \bibinfo {pages} {171} (\bibinfo {year}
  {2004})}\BibitemShut {NoStop}%
\bibitem [{\citenamefont {Li}\ \emph {et~al.}(2012)\citenamefont {Li},
  \citenamefont {Jing}, \citenamefont {Zhang},\ and\ \citenamefont
  {Long}}]{li2012}%
  \BibitemOpen
  \bibfield  {author} {\bibinfo {author} {\bibfnamefont {Y.}~\bibnamefont
  {Li}}, \bibinfo {author} {\bibfnamefont {C.}~\bibnamefont {Jing}}, \bibinfo
  {author} {\bibfnamefont {L.}~\bibnamefont {Zhang}},\ and\ \bibinfo {author}
  {\bibfnamefont {Y.-T.}\ \bibnamefont {Long}},\ }\bibfield  {title} {\bibinfo
  {title} {Resonance scattering particles as biological nanosensors in vitro
  and in vivo},\ }\href {https://doi.org/10.1039/C1CS15143F} {\bibfield
  {journal} {\bibinfo  {journal} {Chem. Soc. Rev.}\ }\textbf {\bibinfo {volume}
  {41}},\ \bibinfo {pages} {632} (\bibinfo {year} {2012})}\BibitemShut
  {NoStop}%
\bibitem [{\citenamefont {Swierczewska}\ \emph {et~al.}(2012)\citenamefont
  {Swierczewska}, \citenamefont {Liu}, \citenamefont {Lee},\ and\ \citenamefont
  {Chen}}]{swierczewska2012}%
  \BibitemOpen
  \bibfield  {author} {\bibinfo {author} {\bibfnamefont {M.}~\bibnamefont
  {Swierczewska}}, \bibinfo {author} {\bibfnamefont {G.}~\bibnamefont {Liu}},
  \bibinfo {author} {\bibfnamefont {S.}~\bibnamefont {Lee}},\ and\ \bibinfo
  {author} {\bibfnamefont {X.}~\bibnamefont {Chen}},\ }\bibfield  {title}
  {\bibinfo {title} {High-sensitivity nanosensors for biomarker detection},\
  }\href {https://doi.org/10.1039/C1CS15238F} {\bibfield  {journal} {\bibinfo
  {journal} {Chem. Soc. Rev.}\ }\textbf {\bibinfo {volume} {41}},\ \bibinfo
  {pages} {2641} (\bibinfo {year} {2012})}\BibitemShut {NoStop}%
\bibitem [{\citenamefont {Smith}\ \emph {et~al.}(2015)\citenamefont {Smith},
  \citenamefont {Stenger}, \citenamefont {Kristensen}, \citenamefont
  {Mortensen},\ and\ \citenamefont {Bozhevolnyi}}]{smith2015}%
  \BibitemOpen
  \bibfield  {author} {\bibinfo {author} {\bibfnamefont {C.~L.~C.}\
  \bibnamefont {Smith}}, \bibinfo {author} {\bibfnamefont {N.}~\bibnamefont
  {Stenger}}, \bibinfo {author} {\bibfnamefont {A.}~\bibnamefont {Kristensen}},
  \bibinfo {author} {\bibfnamefont {N.~A.}\ \bibnamefont {Mortensen}},\ and\
  \bibinfo {author} {\bibfnamefont {S.~I.}\ \bibnamefont {Bozhevolnyi}},\
  }\bibfield  {title} {\bibinfo {title} {Gap and channeled plasmons in tapered
  grooves: a review},\ }\href {https://doi.org/10.1039/C5NR01282A} {\bibfield
  {journal} {\bibinfo  {journal} {Nanoscale}\ }\textbf {\bibinfo {volume}
  {7}},\ \bibinfo {pages} {9355} (\bibinfo {year} {2015})}\BibitemShut
  {NoStop}%
\bibitem [{\citenamefont {Chavda}\ \emph {et~al.}(2023)\citenamefont {Chavda},
  \citenamefont {Balar}, \citenamefont {Nalla}, \citenamefont {Bezbaruah},
  \citenamefont {Gogoi}, \citenamefont {Gajula}, \citenamefont {Peng},
  \citenamefont {Meena}, \citenamefont {Conde},\ and\ \citenamefont
  {Prasad}}]{chavda2023}%
  \BibitemOpen
  \bibfield  {author} {\bibinfo {author} {\bibfnamefont {V.~P.}\ \bibnamefont
  {Chavda}}, \bibinfo {author} {\bibfnamefont {P.~C.}\ \bibnamefont {Balar}},
  \bibinfo {author} {\bibfnamefont {L.~V.}\ \bibnamefont {Nalla}}, \bibinfo
  {author} {\bibfnamefont {R.}~\bibnamefont {Bezbaruah}}, \bibinfo {author}
  {\bibfnamefont {N.~R.}\ \bibnamefont {Gogoi}}, \bibinfo {author}
  {\bibfnamefont {S.~N.~R.}\ \bibnamefont {Gajula}}, \bibinfo {author}
  {\bibfnamefont {B.}~\bibnamefont {Peng}}, \bibinfo {author} {\bibfnamefont
  {A.~S.}\ \bibnamefont {Meena}}, \bibinfo {author} {\bibfnamefont
  {J.}~\bibnamefont {Conde}},\ and\ \bibinfo {author} {\bibfnamefont
  {R.}~\bibnamefont {Prasad}},\ }\bibfield  {title} {\bibinfo {title}
  {Conjugated nanoparticles for solid tumor theranostics: Unraveling the
  interplay of known and unknown factors},\ }\href
  {https://doi.org/10.1021/acsomega.3c05069} {\bibfield  {journal} {\bibinfo
  {journal} {ACS Omega}\ }\textbf {\bibinfo {volume} {8}},\ \bibinfo {pages}
  {37654} (\bibinfo {year} {2023})}\BibitemShut {NoStop}%
\bibitem [{\citenamefont {Lewis}\ and\ \citenamefont
  {Berkelbach}(2019)}]{lewis2019}%
  \BibitemOpen
  \bibfield  {author} {\bibinfo {author} {\bibfnamefont {A.~M.}\ \bibnamefont
  {Lewis}}\ and\ \bibinfo {author} {\bibfnamefont {T.~C.}\ \bibnamefont
  {Berkelbach}},\ }\bibfield  {title} {\bibinfo {title} {Ab initio lifetime and
  concomitant double-excitation character of plasmons at metallic densities},\
  }\href {https://doi.org/10.1103/PhysRevLett.122.226402} {\bibfield  {journal}
  {\bibinfo  {journal} {Phys. Rev. Lett.}\ }\textbf {\bibinfo {volume} {122}},\
  \bibinfo {pages} {226402} (\bibinfo {year} {2019})}\BibitemShut {NoStop}%
\bibitem [{\citenamefont {Li}\ \emph {et~al.}(2023)\citenamefont {Li},
  \citenamefont {Shi}, \citenamefont {Lin},\ and\ \citenamefont
  {Ren}}]{li2023}%
  \BibitemOpen
  \bibfield  {author} {\bibinfo {author} {\bibfnamefont {P.}~\bibnamefont
  {Li}}, \bibinfo {author} {\bibfnamefont {R.}~\bibnamefont {Shi}}, \bibinfo
  {author} {\bibfnamefont {P.}~\bibnamefont {Lin}},\ and\ \bibinfo {author}
  {\bibfnamefont {X.}~\bibnamefont {Ren}},\ }\bibfield  {title} {\bibinfo
  {title} {First-principles calculations of plasmon excitations in graphene,
  silicene, and germanene},\ }\href
  {https://doi.org/10.1103/PhysRevB.107.035433} {\bibfield  {journal} {\bibinfo
   {journal} {Phys. Rev. B}\ }\textbf {\bibinfo {volume} {107}},\ \bibinfo
  {pages} {035433} (\bibinfo {year} {2023})}\BibitemShut {NoStop}%
\bibitem [{\citenamefont {Scott}\ and\ \citenamefont
  {Booth}(2024)}]{scott2024}%
  \BibitemOpen
  \bibfield  {author} {\bibinfo {author} {\bibfnamefont {C.~J.~C.}\
  \bibnamefont {Scott}}\ and\ \bibinfo {author} {\bibfnamefont {G.~H.}\
  \bibnamefont {Booth}},\ }\bibfield  {title} {\bibinfo {title} {Rigorous
  screened interactions for realistic correlated electron systems},\ }\href
  {https://doi.org/10.1103/PhysRevLett.132.076401} {\bibfield  {journal}
  {\bibinfo  {journal} {Phys. Rev. Lett.}\ }\textbf {\bibinfo {volume} {132}},\
  \bibinfo {pages} {076401} (\bibinfo {year} {2024})}\BibitemShut {NoStop}%
\bibitem [{\citenamefont {van Loon}\ \emph
  {et~al.}(2014{\natexlab{a}})\citenamefont {van Loon}, \citenamefont
  {Hafermann}, \citenamefont {Lichtenstein}, \citenamefont {Rubtsov},\ and\
  \citenamefont {Katsnelson}}]{vanloon2014b}%
  \BibitemOpen
  \bibfield  {author} {\bibinfo {author} {\bibfnamefont {E.~G. C.~P.}\
  \bibnamefont {van Loon}}, \bibinfo {author} {\bibfnamefont {H.}~\bibnamefont
  {Hafermann}}, \bibinfo {author} {\bibfnamefont {A.~I.}\ \bibnamefont
  {Lichtenstein}}, \bibinfo {author} {\bibfnamefont {A.~N.}\ \bibnamefont
  {Rubtsov}},\ and\ \bibinfo {author} {\bibfnamefont {M.~I.}\ \bibnamefont
  {Katsnelson}},\ }\bibfield  {title} {\bibinfo {title} {Plasmons in strongly
  correlated systems: Spectral weight transfer and renormalized dispersion},\
  }\href {https://doi.org/10.1103/PhysRevLett.113.246407} {\bibfield  {journal}
  {\bibinfo  {journal} {Phys. Rev. Lett.}\ }\textbf {\bibinfo {volume} {113}},\
  \bibinfo {pages} {246407} (\bibinfo {year} {2014}{\natexlab{a}})}\BibitemShut
  {NoStop}%
\bibitem [{\citenamefont {Hafermann}\ \emph {et~al.}(2014)\citenamefont
  {Hafermann}, \citenamefont {van Loon}, \citenamefont {Katsnelson},
  \citenamefont {Lichtenstein},\ and\ \citenamefont
  {Parcollet}}]{hafermann2014}%
  \BibitemOpen
  \bibfield  {author} {\bibinfo {author} {\bibfnamefont {H.}~\bibnamefont
  {Hafermann}}, \bibinfo {author} {\bibfnamefont {E.~G. C.~P.}\ \bibnamefont
  {van Loon}}, \bibinfo {author} {\bibfnamefont {M.~I.}\ \bibnamefont
  {Katsnelson}}, \bibinfo {author} {\bibfnamefont {A.~I.}\ \bibnamefont
  {Lichtenstein}},\ and\ \bibinfo {author} {\bibfnamefont {O.}~\bibnamefont
  {Parcollet}},\ }\bibfield  {title} {\bibinfo {title} {Collective charge
  excitations of strongly correlated electrons, vertex corrections, and gauge
  invariance},\ }\href {https://doi.org/10.1103/PhysRevB.90.235105} {\bibfield
  {journal} {\bibinfo  {journal} {Phys. Rev. B}\ }\textbf {\bibinfo {volume}
  {90}},\ \bibinfo {pages} {235105} (\bibinfo {year} {2014})}\BibitemShut
  {NoStop}%
\bibitem [{\citenamefont {Guzzo}\ \emph {et~al.}(2011)\citenamefont {Guzzo},
  \citenamefont {Lani}, \citenamefont {Sottile}, \citenamefont {Romaniello},
  \citenamefont {Gatti}, \citenamefont {Kas}, \citenamefont {Rehr},
  \citenamefont {Silly}, \citenamefont {Sirotti},\ and\ \citenamefont
  {Reining}}]{guzzo2011}%
  \BibitemOpen
  \bibfield  {author} {\bibinfo {author} {\bibfnamefont {M.}~\bibnamefont
  {Guzzo}}, \bibinfo {author} {\bibfnamefont {G.}~\bibnamefont {Lani}},
  \bibinfo {author} {\bibfnamefont {F.}~\bibnamefont {Sottile}}, \bibinfo
  {author} {\bibfnamefont {P.}~\bibnamefont {Romaniello}}, \bibinfo {author}
  {\bibfnamefont {M.}~\bibnamefont {Gatti}}, \bibinfo {author} {\bibfnamefont
  {J.~J.}\ \bibnamefont {Kas}}, \bibinfo {author} {\bibfnamefont {J.~J.}\
  \bibnamefont {Rehr}}, \bibinfo {author} {\bibfnamefont {M.~G.}\ \bibnamefont
  {Silly}}, \bibinfo {author} {\bibfnamefont {F.}~\bibnamefont {Sirotti}},\
  and\ \bibinfo {author} {\bibfnamefont {L.}~\bibnamefont {Reining}},\
  }\bibfield  {title} {\bibinfo {title} {Valence electron photoemission
  spectrum of semiconductors: Ab initio description of multiple satellites},\
  }\href {https://doi.org/10.1103/PhysRevLett.107.166401} {\bibfield  {journal}
  {\bibinfo  {journal} {Phys. Rev. Lett.}\ }\textbf {\bibinfo {volume} {107}},\
  \bibinfo {pages} {166401} (\bibinfo {year} {2011})}\BibitemShut {NoStop}%
\bibitem [{\citenamefont {Zhou}\ \emph {et~al.}(2018)\citenamefont {Zhou},
  \citenamefont {Gatti}, \citenamefont {Kas}, \citenamefont {Rehr},\ and\
  \citenamefont {Reining}}]{zhou2018}%
  \BibitemOpen
  \bibfield  {author} {\bibinfo {author} {\bibfnamefont {J.~S.}\ \bibnamefont
  {Zhou}}, \bibinfo {author} {\bibfnamefont {M.}~\bibnamefont {Gatti}},
  \bibinfo {author} {\bibfnamefont {J.~J.}\ \bibnamefont {Kas}}, \bibinfo
  {author} {\bibfnamefont {J.~J.}\ \bibnamefont {Rehr}},\ and\ \bibinfo
  {author} {\bibfnamefont {L.}~\bibnamefont {Reining}},\ }\bibfield  {title}
  {\bibinfo {title} {Cumulant {G}reen's function calculations of plasmon
  satellites in bulk sodium: Influence of screening and the crystal
  environment},\ }\href {https://doi.org/10.1103/PhysRevB.97.035137} {\bibfield
   {journal} {\bibinfo  {journal} {Phys. Rev. B}\ }\textbf {\bibinfo {volume}
  {97}},\ \bibinfo {pages} {035137} (\bibinfo {year} {2018})}\BibitemShut
  {NoStop}%
\bibitem [{\citenamefont {Cudazzo}\ and\ \citenamefont
  {Reining}(2020)}]{cudazzo2020}%
  \BibitemOpen
  \bibfield  {author} {\bibinfo {author} {\bibfnamefont {P.}~\bibnamefont
  {Cudazzo}}\ and\ \bibinfo {author} {\bibfnamefont {L.}~\bibnamefont
  {Reining}},\ }\bibfield  {title} {\bibinfo {title} {Correlation satellites in
  optical and loss spectra},\ }\href
  {https://doi.org/10.1103/PhysRevResearch.2.012032} {\bibfield  {journal}
  {\bibinfo  {journal} {Phys. Rev. Res.}\ }\textbf {\bibinfo {volume} {2}},\
  \bibinfo {pages} {012032(R)} (\bibinfo {year} {2020})}\BibitemShut {NoStop}%
\bibitem [{\citenamefont {Zinni}\ \emph {et~al.}(2023)\citenamefont {Zinni},
  \citenamefont {Bejas}, \citenamefont {Yamase},\ and\ \citenamefont
  {Greco}}]{zinni2023}%
  \BibitemOpen
  \bibfield  {author} {\bibinfo {author} {\bibfnamefont {L.}~\bibnamefont
  {Zinni}}, \bibinfo {author} {\bibfnamefont {M.}~\bibnamefont {Bejas}},
  \bibinfo {author} {\bibfnamefont {H.}~\bibnamefont {Yamase}},\ and\ \bibinfo
  {author} {\bibfnamefont {A.}~\bibnamefont {Greco}},\ }\bibfield  {title}
  {\bibinfo {title} {Low-energy plasmon excitations in infinite-layer
  nickelates},\ }\href {https://doi.org/10.1103/PhysRevB.107.014503} {\bibfield
   {journal} {\bibinfo  {journal} {Phys. Rev. B}\ }\textbf {\bibinfo {volume}
  {107}},\ \bibinfo {pages} {014503} (\bibinfo {year} {2023})}\BibitemShut
  {NoStop}%
\bibitem [{\citenamefont {Kotliar}\ and\ \citenamefont
  {Ruckenstein}(1986)}]{kotliar1986}%
  \BibitemOpen
  \bibfield  {author} {\bibinfo {author} {\bibfnamefont {G.}~\bibnamefont
  {Kotliar}}\ and\ \bibinfo {author} {\bibfnamefont {A.~E.}\ \bibnamefont
  {Ruckenstein}},\ }\bibfield  {title} {\bibinfo {title} {New functional
  integral approach to strongly correlated {F}ermi systems: The {G}utzwiller
  approximation as a saddle point},\ }\href
  {https://doi.org/10.1103/PhysRevLett.57.1362} {\bibfield  {journal} {\bibinfo
   {journal} {Phys. Rev. Lett.}\ }\textbf {\bibinfo {volume} {57}},\ \bibinfo
  {pages} {1362} (\bibinfo {year} {1986})}\BibitemShut {NoStop}%
\bibitem [{\citenamefont {Deeg}\ \emph {et~al.}(1993)\citenamefont {Deeg},
  \citenamefont {Fehske},\ and\ \citenamefont {B\"uttner}}]{deeg1993}%
  \BibitemOpen
  \bibfield  {author} {\bibinfo {author} {\bibfnamefont {M.}~\bibnamefont
  {Deeg}}, \bibinfo {author} {\bibfnamefont {H.}~\bibnamefont {Fehske}},\ and\
  \bibinfo {author} {\bibfnamefont {H.}~\bibnamefont {B\"uttner}},\ }\bibfield
  {title} {\bibinfo {title} {Slave-boson phase diagram of the two-dimensional
  extended {H}ubbard model: Influence of electron-phonon coupling},\ }\href
  {https://doi.org/10.1007/BF01316705} {\bibfield  {journal} {\bibinfo
  {journal} {Z. Phys. B}\ }\textbf {\bibinfo {volume} {91}} (\bibinfo {year}
  {1993})}\BibitemShut {NoStop}%
\bibitem [{\citenamefont {Lhoutellier}\ \emph {et~al.}(2015)\citenamefont
  {Lhoutellier}, \citenamefont {Fr\'esard},\ and\ \citenamefont
  {Ole\ifmmode~\acute{s}\else \'{s}\fi{}}}]{lhoutellier2015}%
  \BibitemOpen
  \bibfield  {author} {\bibinfo {author} {\bibfnamefont {G.}~\bibnamefont
  {Lhoutellier}}, \bibinfo {author} {\bibfnamefont {R.}~\bibnamefont
  {Fr\'esard}},\ and\ \bibinfo {author} {\bibfnamefont {A.~M.}\ \bibnamefont
  {Ole\ifmmode~\acute{s}\else \'{s}\fi{}}},\ }\bibfield  {title} {\bibinfo
  {title} {Fermi-liquid {L}andau parameters for a nondegenerate band: Spin and
  charge instabilities in the extended {H}ubbard model},\ }\href
  {https://doi.org/10.1103/PhysRevB.91.224410} {\bibfield  {journal} {\bibinfo
  {journal} {Phys. Rev. B}\ }\textbf {\bibinfo {volume} {91}},\ \bibinfo
  {pages} {224410} (\bibinfo {year} {2015})}\BibitemShut {NoStop}%
\bibitem [{\citenamefont {Riegler}\ \emph {et~al.}(2023)\citenamefont
  {Riegler}, \citenamefont {Seufert}, \citenamefont {da~Silva~Neto},
  \citenamefont {W\"olfle}, \citenamefont {Thomale},\ and\ \citenamefont
  {Klett}}]{riegler2023}%
  \BibitemOpen
  \bibfield  {author} {\bibinfo {author} {\bibfnamefont {D.}~\bibnamefont
  {Riegler}}, \bibinfo {author} {\bibfnamefont {J.}~\bibnamefont {Seufert}},
  \bibinfo {author} {\bibfnamefont {E.~H.}\ \bibnamefont {da~Silva~Neto}},
  \bibinfo {author} {\bibfnamefont {P.}~\bibnamefont {W\"olfle}}, \bibinfo
  {author} {\bibfnamefont {R.}~\bibnamefont {Thomale}},\ and\ \bibinfo {author}
  {\bibfnamefont {M.}~\bibnamefont {Klett}},\ }\bibfield  {title} {\bibinfo
  {title} {Interplay of spin and charge order in the electron-doped cuprates},\
  }\href {https://doi.org/10.1103/PhysRevB.108.195141} {\bibfield  {journal}
  {\bibinfo  {journal} {Phys. Rev. B}\ }\textbf {\bibinfo {volume} {108}},\
  \bibinfo {pages} {195141} (\bibinfo {year} {2023})}\BibitemShut {NoStop}%
\bibitem [{\citenamefont {Philoxene}\ \emph {et~al.}(2022)\citenamefont
  {Philoxene}, \citenamefont {Dao},\ and\ \citenamefont
  {Fr\'esard}}]{philoxene2022}%
  \BibitemOpen
  \bibfield  {author} {\bibinfo {author} {\bibfnamefont {L.}~\bibnamefont
  {Philoxene}}, \bibinfo {author} {\bibfnamefont {V.~H.}\ \bibnamefont {Dao}},\
  and\ \bibinfo {author} {\bibfnamefont {R.}~\bibnamefont {Fr\'esard}},\
  }\bibfield  {title} {\bibinfo {title} {Spin and charge modulations of a
  half-filled extended {H}ubbard model},\ }\href
  {https://doi.org/10.1103/PhysRevB.106.235131} {\bibfield  {journal} {\bibinfo
   {journal} {Phys. Rev. B}\ }\textbf {\bibinfo {volume} {106}},\ \bibinfo
  {pages} {235131} (\bibinfo {year} {2022})}\BibitemShut {NoStop}%
\bibitem [{Note1()}]{Note1}%
  \BibitemOpen
  \bibinfo {note} {In particular, the paramagnetic saddle-point approximation
  of the KRSB representation is equivalent to the Gutzwiller approximation,
  thereby describing the MIT at half filling. On the cubic lattice, the MIT
  critical coupling $U_c=1.33W$~\cite {lhoutellier2015}, where $W$ is the bare
  bandwidth, compares favorably with the DMFT result of $U_c=1.17W$~\cite
  {zitko2009}.}\BibitemShut {Stop}%
\bibitem [{\citenamefont {Dao}\ and\ \citenamefont
  {Fr\'esard}(2017)}]{dao2017}%
  \BibitemOpen
  \bibfield  {author} {\bibinfo {author} {\bibfnamefont {V.~H.}\ \bibnamefont
  {Dao}}\ and\ \bibinfo {author} {\bibfnamefont {R.}~\bibnamefont
  {Fr\'esard}},\ }\bibfield  {title} {\bibinfo {title} {Collective modes in the
  paramagnetic phase of the {H}ubbard model},\ }\href
  {https://doi.org/10.1103/PhysRevB.95.165127} {\bibfield  {journal} {\bibinfo
  {journal} {Phys. Rev. B}\ }\textbf {\bibinfo {volume} {95}},\ \bibinfo
  {pages} {165127} (\bibinfo {year} {2017})}\BibitemShut {NoStop}%
\bibitem [{\citenamefont {Aichhorn}\ \emph {et~al.}(2004)\citenamefont
  {Aichhorn}, \citenamefont {Evertz}, \citenamefont {von~der Linden},\ and\
  \citenamefont {Potthoff}}]{aichhorn2004}%
  \BibitemOpen
  \bibfield  {author} {\bibinfo {author} {\bibfnamefont {M.}~\bibnamefont
  {Aichhorn}}, \bibinfo {author} {\bibfnamefont {H.~G.}\ \bibnamefont
  {Evertz}}, \bibinfo {author} {\bibfnamefont {W.}~\bibnamefont {von~der
  Linden}},\ and\ \bibinfo {author} {\bibfnamefont {M.}~\bibnamefont
  {Potthoff}},\ }\bibfield  {title} {\bibinfo {title} {Charge ordering in
  extended {H}ubbard models: Variational cluster approach},\ }\href
  {https://doi.org/10.1103/PhysRevB.70.235107} {\bibfield  {journal} {\bibinfo
  {journal} {Phys. Rev. B}\ }\textbf {\bibinfo {volume} {70}},\ \bibinfo
  {pages} {235107} (\bibinfo {year} {2004})}\BibitemShut {NoStop}%
\bibitem [{\citenamefont {Davoudi}\ and\ \citenamefont
  {Tremblay}(2006)}]{davoudi2006}%
  \BibitemOpen
  \bibfield  {author} {\bibinfo {author} {\bibfnamefont {B.}~\bibnamefont
  {Davoudi}}\ and\ \bibinfo {author} {\bibfnamefont {A.-M.~S.}\ \bibnamefont
  {Tremblay}},\ }\bibfield  {title} {\bibinfo {title} {Nearest-neighbor
  repulsion and competing charge and spin order in the extended {H}ubbard
  model},\ }\href {https://doi.org/10.1103/PhysRevB.74.035113} {\bibfield
  {journal} {\bibinfo  {journal} {Phys. Rev. B}\ }\textbf {\bibinfo {volume}
  {74}},\ \bibinfo {pages} {035113} (\bibinfo {year} {2006})}\BibitemShut
  {NoStop}%
\bibitem [{\citenamefont {Kagan}\ \emph {et~al.}(2011)\citenamefont {Kagan},
  \citenamefont {Efremov}, \citenamefont {Marienko},\ and\ \citenamefont
  {Val’kov}}]{kagan2011}%
  \BibitemOpen
  \bibfield  {author} {\bibinfo {author} {\bibfnamefont {M.~Y.}\ \bibnamefont
  {Kagan}}, \bibinfo {author} {\bibfnamefont {D.~V.}\ \bibnamefont {Efremov}},
  \bibinfo {author} {\bibfnamefont {M.~S.}\ \bibnamefont {Marienko}},\ and\
  \bibinfo {author} {\bibfnamefont {V.~V.}\ \bibnamefont {Val’kov}},\
  }\bibfield  {title} {\bibinfo {title} {Triplet p-wave superconductivity in
  the low-density extended {H}ubbard model with {C}oulomg repulsion},\ }\href
  {https://doi.org/10.1134/S0021364011120083} {\bibfield  {journal} {\bibinfo
  {journal} {JETP Lett.}\ }\textbf {\bibinfo {volume} {93}},\ \bibinfo {pages}
  {725} (\bibinfo {year} {2011})}\BibitemShut {NoStop}%
\bibitem [{\citenamefont {Ayral}\ \emph {et~al.}(2013)\citenamefont {Ayral},
  \citenamefont {Biermann},\ and\ \citenamefont {Werner}}]{ayral2013}%
  \BibitemOpen
  \bibfield  {author} {\bibinfo {author} {\bibfnamefont {T.}~\bibnamefont
  {Ayral}}, \bibinfo {author} {\bibfnamefont {S.}~\bibnamefont {Biermann}},\
  and\ \bibinfo {author} {\bibfnamefont {P.}~\bibnamefont {Werner}},\
  }\bibfield  {title} {\bibinfo {title} {Screening and nonlocal correlations in
  the extended {H}ubbard model from self-consistent combined {$GW$} and
  dynamical mean field theory},\ }\href
  {https://doi.org/10.1103/PhysRevB.87.125149} {\bibfield  {journal} {\bibinfo
  {journal} {Phys. Rev. B}\ }\textbf {\bibinfo {volume} {87}},\ \bibinfo
  {pages} {125149} (\bibinfo {year} {2013})}\BibitemShut {NoStop}%
\bibitem [{\citenamefont {van Loon}\ \emph
  {et~al.}(2014{\natexlab{b}})\citenamefont {van Loon}, \citenamefont
  {Lichtenstein}, \citenamefont {Katsnelson}, \citenamefont {Parcollet},\ and\
  \citenamefont {Hafermann}}]{vanloon2014}%
  \BibitemOpen
  \bibfield  {author} {\bibinfo {author} {\bibfnamefont {E.~G. C.~P.}\
  \bibnamefont {van Loon}}, \bibinfo {author} {\bibfnamefont {A.~I.}\
  \bibnamefont {Lichtenstein}}, \bibinfo {author} {\bibfnamefont {M.~I.}\
  \bibnamefont {Katsnelson}}, \bibinfo {author} {\bibfnamefont
  {O.}~\bibnamefont {Parcollet}},\ and\ \bibinfo {author} {\bibfnamefont
  {H.}~\bibnamefont {Hafermann}},\ }\bibfield  {title} {\bibinfo {title}
  {Beyond extended dynamical mean-field theory: Dual boson approach to the
  two-dimensional extended {H}ubbard model},\ }\href
  {https://doi.org/10.1103/PhysRevB.90.235135} {\bibfield  {journal} {\bibinfo
  {journal} {Phys. Rev. B}\ }\textbf {\bibinfo {volume} {90}},\ \bibinfo
  {pages} {235135} (\bibinfo {year} {2014}{\natexlab{b}})}\BibitemShut
  {NoStop}%
\bibitem [{\citenamefont {Kapcia}\ \emph {et~al.}(2017)\citenamefont {Kapcia},
  \citenamefont {Robaszkiewicz}, \citenamefont {Capone},\ and\ \citenamefont
  {Amaricci}}]{kapcia2017}%
  \BibitemOpen
  \bibfield  {author} {\bibinfo {author} {\bibfnamefont {K.~J.}\ \bibnamefont
  {Kapcia}}, \bibinfo {author} {\bibfnamefont {S.}~\bibnamefont
  {Robaszkiewicz}}, \bibinfo {author} {\bibfnamefont {M.}~\bibnamefont
  {Capone}},\ and\ \bibinfo {author} {\bibfnamefont {A.}~\bibnamefont
  {Amaricci}},\ }\bibfield  {title} {\bibinfo {title} {Doping-driven
  metal-insulator transitions and charge orderings in the extended {H}ubbard
  model},\ }\href {https://doi.org/10.1103/PhysRevB.95.125112} {\bibfield
  {journal} {\bibinfo  {journal} {Phys. Rev. B}\ }\textbf {\bibinfo {volume}
  {95}},\ \bibinfo {pages} {125112} (\bibinfo {year} {2017})}\BibitemShut
  {NoStop}%
\bibitem [{\citenamefont {Sch\"uler}\ \emph {et~al.}(2018)\citenamefont
  {Sch\"uler}, \citenamefont {van Loon}, \citenamefont {Katsnelson},\ and\
  \citenamefont {Wehling}}]{schuller2018}%
  \BibitemOpen
  \bibfield  {author} {\bibinfo {author} {\bibfnamefont {M.}~\bibnamefont
  {Sch\"uler}}, \bibinfo {author} {\bibfnamefont {E.~G. C.~P.}\ \bibnamefont
  {van Loon}}, \bibinfo {author} {\bibfnamefont {M.~I.}\ \bibnamefont
  {Katsnelson}},\ and\ \bibinfo {author} {\bibfnamefont {T.~O.}\ \bibnamefont
  {Wehling}},\ }\bibfield  {title} {\bibinfo {title} {First-order
  metal-insulator transitions in the extended {H}ubbard model due to
  self-consistent screening of the effective interaction},\ }\href
  {https://doi.org/10.1103/PhysRevB.97.165135} {\bibfield  {journal} {\bibinfo
  {journal} {Phys. Rev. B}\ }\textbf {\bibinfo {volume} {97}},\ \bibinfo
  {pages} {165135} (\bibinfo {year} {2018})}\BibitemShut {NoStop}%
\bibitem [{\citenamefont {Terletska}\ \emph {et~al.}(2021)\citenamefont
  {Terletska}, \citenamefont {Iskakov}, \citenamefont {Maier},\ and\
  \citenamefont {Gull}}]{terletska2021}%
  \BibitemOpen
  \bibfield  {author} {\bibinfo {author} {\bibfnamefont {H.}~\bibnamefont
  {Terletska}}, \bibinfo {author} {\bibfnamefont {S.}~\bibnamefont {Iskakov}},
  \bibinfo {author} {\bibfnamefont {T.}~\bibnamefont {Maier}},\ and\ \bibinfo
  {author} {\bibfnamefont {E.}~\bibnamefont {Gull}},\ }\bibfield  {title}
  {\bibinfo {title} {Dynamical cluster approximation study of electron
  localization in the extended {H}ubbard model},\ }\href
  {https://doi.org/10.1103/PhysRevB.104.085129} {\bibfield  {journal} {\bibinfo
   {journal} {Phys. Rev. B}\ }\textbf {\bibinfo {volume} {104}},\ \bibinfo
  {pages} {085129} (\bibinfo {year} {2021})}\BibitemShut {NoStop}%
\bibitem [{\citenamefont {Roig}\ \emph {et~al.}(2022)\citenamefont {Roig},
  \citenamefont {R\o{}mer}, \citenamefont {Hirschfeld},\ and\ \citenamefont
  {Andersen}}]{roig2022}%
  \BibitemOpen
  \bibfield  {author} {\bibinfo {author} {\bibfnamefont {M.}~\bibnamefont
  {Roig}}, \bibinfo {author} {\bibfnamefont {A.~T.}\ \bibnamefont {R\o{}mer}},
  \bibinfo {author} {\bibfnamefont {P.~J.}\ \bibnamefont {Hirschfeld}},\ and\
  \bibinfo {author} {\bibfnamefont {B.~M.}\ \bibnamefont {Andersen}},\
  }\bibfield  {title} {\bibinfo {title} {Revisiting superconductivity in the
  extended one-band {H}ubbard model: Pairing via spin and charge
  fluctuations},\ }\href {https://doi.org/10.1103/PhysRevB.106.214530}
  {\bibfield  {journal} {\bibinfo  {journal} {Phys. Rev. B}\ }\textbf {\bibinfo
  {volume} {106}},\ \bibinfo {pages} {214530} (\bibinfo {year}
  {2022})}\BibitemShut {NoStop}%
\bibitem [{\citenamefont {Linn\'er}\ \emph {et~al.}(2023)\citenamefont
  {Linn\'er}, \citenamefont {Lichtenstein}, \citenamefont {Biermann},\ and\
  \citenamefont {Stepanov}}]{linner2023}%
  \BibitemOpen
  \bibfield  {author} {\bibinfo {author} {\bibfnamefont {E.}~\bibnamefont
  {Linn\'er}}, \bibinfo {author} {\bibfnamefont {A.~I.}\ \bibnamefont
  {Lichtenstein}}, \bibinfo {author} {\bibfnamefont {S.}~\bibnamefont
  {Biermann}},\ and\ \bibinfo {author} {\bibfnamefont {E.~A.}\ \bibnamefont
  {Stepanov}},\ }\bibfield  {title} {\bibinfo {title} {Multichannel fluctuating
  field approach to competing instabilities in interacting electronic
  systems},\ }\href {https://doi.org/10.1103/PhysRevB.108.035143} {\bibfield
  {journal} {\bibinfo  {journal} {Phys. Rev. B}\ }\textbf {\bibinfo {volume}
  {108}},\ \bibinfo {pages} {035143} (\bibinfo {year} {2023})}\BibitemShut
  {NoStop}%
\bibitem [{\citenamefont {Kundu}\ and\ \citenamefont
  {Sénéchal}(2024)}]{kundu2023}%
  \BibitemOpen
  \bibfield  {author} {\bibinfo {author} {\bibfnamefont {S.}~\bibnamefont
  {Kundu}}\ and\ \bibinfo {author} {\bibfnamefont {D.}~\bibnamefont
  {Sénéchal}},\ }\bibfield  {title} {\bibinfo {title} {{CDMFT+HFD: An
  extension of dynamical mean field theory for nonlocal interactions applied to
  the single band extended Hubbard model}},\ }\href
  {https://doi.org/10.21468/SciPostPhysCore.7.2.033} {\bibfield  {journal}
  {\bibinfo  {journal} {SciPost Phys. Core}\ }\textbf {\bibinfo {volume} {7}},\
  \bibinfo {pages} {033} (\bibinfo {year} {2024})}\BibitemShut {NoStop}%
\bibitem [{\citenamefont {Fr\'esard}\ and\ \citenamefont
  {Kopp}(2001)}]{fresard2001}%
  \BibitemOpen
  \bibfield  {author} {\bibinfo {author} {\bibfnamefont {R.}~\bibnamefont
  {Fr\'esard}}\ and\ \bibinfo {author} {\bibfnamefont {T.}~\bibnamefont
  {Kopp}},\ }\bibfield  {title} {\bibinfo {title} {Slave bosons in radial
  gauge: the correct functional integral representation and inclusion of
  non-local interactions},\ }\href
  {https://doi.org/https://doi.org/10.1016/S0550-3213(00)00657-X} {\bibfield
  {journal} {\bibinfo  {journal} {Nucl. Phys. B}\ }\textbf {\bibinfo {volume}
  {594}},\ \bibinfo {pages} {769} (\bibinfo {year} {2001})}\BibitemShut
  {NoStop}%
\bibitem [{\citenamefont {Dao}\ and\ \citenamefont
  {Fr\'esard}(2020)}]{dao2020}%
  \BibitemOpen
  \bibfield  {author} {\bibinfo {author} {\bibfnamefont {V.~H.}\ \bibnamefont
  {Dao}}\ and\ \bibinfo {author} {\bibfnamefont {R.}~\bibnamefont
  {Fr\'esard}},\ }\bibfield  {title} {\bibinfo {title} {Combining complex and
  radial slave boson fields within the {K}otliar-{R}uckenstein representation
  of correlated impurities},\ }\href
  {https://doi.org/https://doi.org/10.1002/andp.201900491} {\bibfield
  {journal} {\bibinfo  {journal} {Ann. Phys. (Berlin)}\ }\textbf {\bibinfo
  {volume} {532}},\ \bibinfo {pages} {1900491} (\bibinfo {year}
  {2020})}\BibitemShut {NoStop}%
\bibitem [{\citenamefont {Dao}\ and\ \citenamefont {Frésard}(2024)}]{dao2024}%
  \BibitemOpen
  \bibfield  {author} {\bibinfo {author} {\bibfnamefont {V.~H.}\ \bibnamefont
  {Dao}}\ and\ \bibinfo {author} {\bibfnamefont {R.}~\bibnamefont {Frésard}},\
  }\bibfield  {title} {\bibinfo {title} {Exact functional integration of radial
  and complex slave-boson fields: Thermodynamics and dynamics of the two-site
  extended Hubbard model},\ }\href
  {https://doi.org/https://doi.org/10.1002/andp.202400029} {\bibfield
  {journal} {\bibinfo  {journal} {Ann. Phys. (Berlin)}\ }\textbf {\bibinfo
  {volume} {536}},\ \bibinfo {pages} {2400029} (\bibinfo {year}
  {2024})}\BibitemShut {NoStop}%
\bibitem [{\citenamefont {Jolicoeur}\ and\ \citenamefont
  {Le~Guillou}(1991)}]{jolicoeur1991}%
  \BibitemOpen
  \bibfield  {author} {\bibinfo {author} {\bibfnamefont {T.}~\bibnamefont
  {Jolicoeur}}\ and\ \bibinfo {author} {\bibfnamefont {J.~C.}\ \bibnamefont
  {Le~Guillou}},\ }\bibfield  {title} {\bibinfo {title} {Fluctuations beyond
  the {G}utzwiller approximation in the slave-boson approach},\ }\href
  {https://doi.org/10.1103/PhysRevB.44.2403} {\bibfield  {journal} {\bibinfo
  {journal} {Phys. Rev. B}\ }\textbf {\bibinfo {volume} {44}},\ \bibinfo
  {pages} {2403} (\bibinfo {year} {1991})}\BibitemShut {NoStop}%
\bibitem [{\citenamefont {Fr\'esard}\ and\ \citenamefont
  {W{\"o}lfle}(1992)}]{fresard1992}%
  \BibitemOpen
  \bibfield  {author} {\bibinfo {author} {\bibfnamefont {R.}~\bibnamefont
  {Fr\'esard}}\ and\ \bibinfo {author} {\bibfnamefont {P.}~\bibnamefont
  {W{\"o}lfle}},\ }\bibfield  {title} {\bibinfo {title} {Unified slave boson
  representation of spin and charge degrees of freedom for strongly correlated
  {F}ermi systems},\ }\href {https://doi.org/10.1142/S0217979292000414}
  {\bibfield  {journal} {\bibinfo  {journal} {Int. J. Mod. Phys. B}\ }\textbf
  {\bibinfo {volume} {06}},\ \bibinfo {pages} {685} (\bibinfo {year}
  {1992})}\BibitemShut {NoStop}%
\bibitem [{\citenamefont {Bang}\ \emph {et~al.}(1992)\citenamefont {Bang},
  \citenamefont {Castellani}, \citenamefont {Grilli}, \citenamefont {Kotliar},
  \citenamefont {Raimondi},\ and\ \citenamefont {Wang}}]{bang1992}%
  \BibitemOpen
  \bibfield  {author} {\bibinfo {author} {\bibfnamefont {Y.}~\bibnamefont
  {Bang}}, \bibinfo {author} {\bibfnamefont {C.}~\bibnamefont {Castellani}},
  \bibinfo {author} {\bibfnamefont {M.}~\bibnamefont {Grilli}}, \bibinfo
  {author} {\bibfnamefont {G.}~\bibnamefont {Kotliar}}, \bibinfo {author}
  {\bibfnamefont {R.}~\bibnamefont {Raimondi}},\ and\ \bibinfo {author}
  {\bibfnamefont {Z.}~\bibnamefont {Wang}},\ }\bibfield  {title} {\bibinfo
  {title} {Single particle and optical gaps in charge-transfer insulators},\
  }\href {https://doi.org/10.1142/S0217979292000311} {\bibfield  {journal}
  {\bibinfo  {journal} {Int. J. Mod. Phys. B}\ }\textbf {\bibinfo {volume}
  {06}},\ \bibinfo {pages} {531} (\bibinfo {year} {1992})}\BibitemShut
  {NoStop}%
\bibitem [{\citenamefont {B{\"u}nemann}\ \emph {et~al.}(2013)\citenamefont
  {B{\"u}nemann}, \citenamefont {Capone}, \citenamefont {Lorenzana},\ and\
  \citenamefont {Seibold}}]{bunemann2013}%
  \BibitemOpen
  \bibfield  {author} {\bibinfo {author} {\bibfnamefont {J.}~\bibnamefont
  {B{\"u}nemann}}, \bibinfo {author} {\bibfnamefont {M.}~\bibnamefont
  {Capone}}, \bibinfo {author} {\bibfnamefont {J.}~\bibnamefont {Lorenzana}},\
  and\ \bibinfo {author} {\bibfnamefont {G.}~\bibnamefont {Seibold}},\
  }\bibfield  {title} {\bibinfo {title} {Linear-response dynamics from the
  time-dependent {G}utzwiller approximation},\ }\href
  {https://doi.org/10.1088/1367-2630/15/5/053050} {\bibfield  {journal}
  {\bibinfo  {journal} {New J. Phys.}\ }\textbf {\bibinfo {volume} {15}},\
  \bibinfo {pages} {053050} (\bibinfo {year} {2013})}\BibitemShut {NoStop}%
\bibitem [{\citenamefont {Noatschk}\ \emph {et~al.}(2020)\citenamefont
  {Noatschk}, \citenamefont {Martens},\ and\ \citenamefont
  {Seibold}}]{noatschk2020}%
  \BibitemOpen
  \bibfield  {author} {\bibinfo {author} {\bibfnamefont {K.}~\bibnamefont
  {Noatschk}}, \bibinfo {author} {\bibfnamefont {C.}~\bibnamefont {Martens}},\
  and\ \bibinfo {author} {\bibfnamefont {G.}~\bibnamefont {Seibold}},\
  }\bibfield  {title} {\bibinfo {title} {Time-dependent {G}utzwiller
  approximation: Theory and applications},\ }\href
  {https://doi.org/10.1007/s10948-019-05406-z} {\bibfield  {journal} {\bibinfo
  {journal} {J. Supercond. Nov. Mag.}\ }\textbf {\bibinfo {volume} {33}},\
  \bibinfo {pages} {2389} (\bibinfo {year} {2020})}\BibitemShut {NoStop}%
\bibitem [{\citenamefont {Vollhardt}\ \emph {et~al.}(1987)\citenamefont
  {Vollhardt}, \citenamefont {W\"olfle},\ and\ \citenamefont
  {Anderson}}]{vollhardt1987}%
  \BibitemOpen
  \bibfield  {author} {\bibinfo {author} {\bibfnamefont {D.}~\bibnamefont
  {Vollhardt}}, \bibinfo {author} {\bibfnamefont {P.}~\bibnamefont
  {W\"olfle}},\ and\ \bibinfo {author} {\bibfnamefont {P.~W.}\ \bibnamefont
  {Anderson}},\ }\bibfield  {title} {\bibinfo {title} {Gutzwiller-{H}ubbard
  lattice-gas model with variable density: Application to normal liquid
  $^{3}\mathrm{He}$},\ }\href {https://doi.org/10.1103/PhysRevB.35.6703}
  {\bibfield  {journal} {\bibinfo  {journal} {Phys. Rev. B}\ }\textbf {\bibinfo
  {volume} {35}},\ \bibinfo {pages} {6703} (\bibinfo {year}
  {1987})}\BibitemShut {NoStop}%
\bibitem [{\citenamefont {Bulla}(1999)}]{bulla1999}%
  \BibitemOpen
  \bibfield  {author} {\bibinfo {author} {\bibfnamefont {R.}~\bibnamefont
  {Bulla}},\ }\bibfield  {title} {\bibinfo {title} {Zero temperature
  metal-insulator transition in the infinite-dimensional hubbard model},\
  }\href {https://doi.org/10.1103/PhysRevLett.83.136} {\bibfield  {journal}
  {\bibinfo  {journal} {Phys. Rev. Lett.}\ }\textbf {\bibinfo {volume} {83}},\
  \bibinfo {pages} {136} (\bibinfo {year} {1999})}\BibitemShut {NoStop}%
\bibitem [{\citenamefont {Zimmermann}\ \emph {et~al.}(1997)\citenamefont
  {Zimmermann}, \citenamefont {Fr\'esard},\ and\ \citenamefont
  {W\"olfle}}]{zimmermann1997}%
  \BibitemOpen
  \bibfield  {author} {\bibinfo {author} {\bibfnamefont {W.}~\bibnamefont
  {Zimmermann}}, \bibinfo {author} {\bibfnamefont {R.}~\bibnamefont
  {Fr\'esard}},\ and\ \bibinfo {author} {\bibfnamefont {P.}~\bibnamefont
  {W\"olfle}},\ }\bibfield  {title} {\bibinfo {title} {Spin and charge
  structure factor of the two-dimensional {H}ubbard model},\ }\href
  {https://doi.org/10.1103/PhysRevB.56.10097} {\bibfield  {journal} {\bibinfo
  {journal} {Phys. Rev. B}\ }\textbf {\bibinfo {volume} {56}},\ \bibinfo
  {pages} {10097} (\bibinfo {year} {1997})}\BibitemShut {NoStop}%
\bibitem [{Note2()}]{Note2}%
  \BibitemOpen
  \bibinfo {note} {Note that, around $\Gamma $, the deviations of
  $f^s({\protect \bf q},\omega )$ from $\protect \frac {U}{2}+V_{\protect \bf
  q}$ appear to vanish. This is, however, a graphical artifact which results
  from the divergence of $V_{\protect \bf q}$ at $\Gamma $, as the data is
  expressed in percents of $\protect \frac {U}{2}+V_{\protect \bf
  q}$.}\BibitemShut {Stop}%
\bibitem [{\citenamefont {\ifmmode~\check{Z}\else \v{Z}\fi{}itko}\ \emph
  {et~al.}(2009)\citenamefont {\ifmmode~\check{Z}\else \v{Z}\fi{}itko},
  \citenamefont {Bon\ifmmode~\check{c}\else \v{c}\fi{}a},\ and\ \citenamefont
  {Pruschke}}]{zitko2009}%
  \BibitemOpen
  \bibfield  {author} {\bibinfo {author} {\bibfnamefont {R.}~\bibnamefont
  {\ifmmode~\check{Z}\else \v{Z}\fi{}itko}}, \bibinfo {author} {\bibfnamefont
  {J.}~\bibnamefont {Bon\ifmmode~\check{c}\else \v{c}\fi{}a}},\ and\ \bibinfo
  {author} {\bibfnamefont {T.}~\bibnamefont {Pruschke}},\ }\bibfield  {title}
  {\bibinfo {title} {Van hove singularities in the paramagnetic phase of the
  hubbard model: Dmft study},\ }\href
  {https://doi.org/10.1103/PhysRevB.80.245112} {\bibfield  {journal} {\bibinfo
  {journal} {Phys. Rev. B}\ }\textbf {\bibinfo {volume} {80}},\ \bibinfo
  {pages} {245112} (\bibinfo {year} {2009})}\BibitemShut {NoStop}%
\end{thebibliography}
\end{document}